\documentclass[usenatbib]{mnras}
\usepackage{times}
\usepackage[T1]{fontenc}
\usepackage{aecompl}
\usepackage{graphicx,bm,verbatim}
\usepackage{multirow}
\usepackage{amsmath,amssymb}
\usepackage{xcolor,ulem}
\usepackage{psfrag}
\def \cb {\textcolor{blue}}

\definecolor{webgreen}{rgb}{0,.5,0}
\definecolor{webbrown}{rgb}{.6,0,0}
\newcommand{\RefC}[1]{\color{black}{#1}}
\newcommand{\RefCC}[1]{\color{black}{#1}}
\newcommand{\HALF}{\frac{1}{2}}
\newcommand{\pd}[2]{\frac{\partial #1}{\partial #2} }
\renewcommand{\vec}[1]{\bm{#1}}
\newcommand{\hvec}[1]{\bm{\hat{#1}}}
\newcommand{\tens}[1]{\mathsf{#1}}
\newcommand{\DS}{\displaystyle}
\newcommand{\legP}[2]{\mathcal{P}_{#1}\left(#2\right)}
\newcommand{\legPp}[2]{\mathcal{P}^{'}_{#1}(#2)}
\newcommand{\legPpp}[2]{\mathcal{P}^{''}_{#1}(#2)}
\newcommand{\matm}{\mathsf{M}}
\hypersetup{colorlinks=true,
            breaklinks=true,
            plainpages=false,
            bookmarksnumbered,
            bookmarksopen=true,
            bookmarksopenlevel=1,
            urlcolor=webbrown,
            linkcolor=blue,
            citecolor=webgreen
            }

\title[RKL-STS for anisotropic diffusion]{Scalable explicit implementation of anisotropic diffusion with Runge-Kutta-Legendre super-time-stepping}
\author[Vaidya, Prasad, Mignone, Sharma and Rickler]
{Bhargav Vaidya$^{1,2}$\thanks{bvaidya@iiti.ac.in, bhargav.vaidya@unito.it}, Deovrat Prasad$^3$, Andrea Mignone$^2$, Prateek Sharma$^3$, Luca Rickler$^2$\\
$^1$Centre of Astronomy, Indian Institute of Technology Indore, Simrol, Khandwa Road, Indore 453552, India \\
$^2$Dipartimento di Fisica Generale, Universita degli Studi di Torino , Via Pietro Giuria 1, 10125 Torino, Italy\\
$^3$Joint Astronomy Programme and Department of Physics, Indian Institute of Science, Bangalore 560012, India\\
}
            
\begin{document}
\maketitle

\begin{abstract}
An important ingredient in numerical modelling of high temperature magnetised astrophysical plasmas is the anisotropic transport of heat along magnetic field lines from higher to lower temperatures.Magnetohydrodynamics (MHD) typically involves solving the hyperbolic set of conservation equations along with the induction equation. Incorporating anisotropic thermal conduction requires to also treat parabolic terms arising from the diffusion operator. 
An explicit treatment of parabolic terms will considerably reduce the simulation time step due to its dependence on the square of the grid resolution ($\Delta x$) for 
stability. Although an implicit scheme relaxes the constraint on stability, 
it is difficult to distribute efficiently on a parallel architecture. 
Treating parabolic terms with accelerated super-time stepping (STS) methods has been discussed in literature 
but these methods suffer from poor accuracy (first order in time) and also have difficult-to-choose tuneable stability parameters. In this work we
highlight a second order (in time) Runge Kutta Legendre (RKL) scheme (first described by \citealt{meyer2012second}) 
that is robust, fast and accurate in treating parabolic terms alongside the hyperbolic conversation laws. We demonstrate its superiority 
over the first order super time stepping schemes  with standard tests and 
astrophysical applications. We also show that explicit conduction is particularly robust in handling saturated thermal conduction. Parallel scaling of  
explicit conduction using RKL scheme is demonstrated up to
more than $10^4$ processors.
\end{abstract}


\section{Introduction}
%
%

Since most baryons in the universe are in a magnetised plasma state, magnetic fields play a crucial role in the dynamics and thermodynamics of astrophysical objects --- ranging from stars and interstellar medium to the intra-cluster and intergalactic medium. Magnetohydrodynamic (MHD) simulations have matured (\citealt{1988ApJ...332..659E,2000JCoPh.161..605T,2001JCoPh.174..614B}) and have contributed to several breakthroughs in our understanding, from accretion to the interstellar medium 
(\citealt{1995ApJ...440..742H,1999ApJ...514L..99K}).
Magnetic fields not only produce forces and stresses in a plasma, they also affect transport properties by predominantly allowing diffusion of heat and momentum along field lines and suppressing transport across them. Anisotropic transport affects 
fundamental properties such as convection/buoyancy in stratified plasmas (\citealt{2000ApJ...534..420B,2008ApJ...673..758Q}), 
and thermal instability and condensation of cold gas out of the hot phase (\citealt{1965ApJ...142..531F,2010ApJ...720..652S}).

While the numerical solution of the ideal MHD equations can be carried out very accurately in highly nonlinear regimes across thousands of processors (\citealt{2011PhRvL.106g5001B,2011ApJ...731...62F}), simulations with anisotropic thermal conduction (and, likewise, similar diffusive processes) have not yet reached the same level of fidelity and scalability.

Taking into account diffusion processes changes the mathematical structure of the underlying system of conservation laws from purely hyperbolic to mixed hyperbolic-parabolic type.
The numerical discretisation of such systems can then pose a more restrictive limitation on the choice of the time step which, for a standard explicit scheme, must scale with the \textit{square} of the grid size rather than with $\Delta x$ alone.
In addition, when saturation effects are considered, thermal conduction itself becomes  a mixed hyperbolic/parabolic operator.
Two approaches are commonly employed to circumvent the stability time step constraint.

In the first one, the parabolic part of the equations is solved using fully- or semi-implicit methods (\citealt{1986nras.book.....P}), allowing for a time step much longer than the explicit one (see e.g., \citealt{2011JCoPh.230.4899S}).
However, scalable implementation of implicit schemes on massively parallel clusters is technically challenging (\citealt{kannan2015accurately}) as it requires solving large sparse matrices.

In the second approach, a particular stability polynomial can be employed to construct an explicit multi-stage time stepping scheme that has extended stability properties.
If $s$ is the number of stages, the length of the super-step that one can take (asymptotically) can be shown to scale as $\sim s^2\Delta t_{\rm p}$, where $\Delta t_{\rm p}$ is the standard explicit parabolic time step.
This technique in general is classified as super-time-stepping (STS). This technique offers therefore an overall speedup $\propto s$ and, being explicit, it can be easily parallelized.
\citealt{ZAMM:ZAMM19800601005,verwer1990convergence} introduced 
Runge-Kutta Chebyshev (RKC-STS) methods. 
A variant of this method was popularized by \citealt{alexiades1996super} (AAG-STS).
More recently, super-time-stepping based on Legendre polynomials, known as  Runge-Kutta Legendre STS (RKL-STS), was proposed by \citet{meyer2012second,meyer2014stabilized}.

In this paper we review and compare the various STS schemes in the context of anisotropic thermal conduction in a MHD plasma.
We show that RKL-STS method is an attractive option for simulating anisotropic diffusion on massively-parallel supercomputers and offers enhanced stability with respect to its predecessors, in particular, AAG-STS.
We note that RKL-STS can also be used to model other isotropic and anisotropic transport processes such as plasma viscosity (e.g., \citealt{2009ApJ...704.1309D}), non-ideal terms in Ohm's law (e.g.,\citealt{OSul2007}) , and cosmic ray diffusion (e.g., \citealt{2009ApJ...699..348S,Pakmor:2016}).
There are also applications in other diverse areas such as image processing (e.g., 
\citealt{weickert1998anisotropic}).

The paper is structured as follows.
In section \ref{sec:gov_eqns} we review the fundamental equations of MHD with thermal conduction.
In Section \ref{sec:num} we review the numerical approaches to anisotropic thermal conduction and also discuss the monotonicity properties in multiple dimensions.
Selected numerical benchmarks are presented in \ref{sec:tests}. The parallel scalability of explicit conduction is demonstrated in \ref{sec:par_scaling} and conclusions are drawn in section \ref{sec:conclusions}. 

\section{Governing Equations}
\label{sec:gov_eqns}
%
%

Here we consider the magnetohydrodynamic (MHD) equations in presence of conduction:
\begin{subequations}
  \begin{align}
  \label{eq:mass}
  \pd{\rho}{t} + \nabla \cdot (\rho \bm{v}) &= 0, \\
  \label{eq:mom}
  \pd{}{t} \left( \rho \vec{v} \right) + \nabla \cdot
    \left(   \rho \bm{v}\vec{v} - \vec{BB} + p_t\tens{I} \right ) &= 0, \\
  \label{eq:energy}
  \pd{E}{t} +  \nabla \cdot\left[ \left( E + p_t\right) \vec{v}
              -  (\vec{B}\cdot\vec{v}) \vec{B} \right] &= - \nabla\cdot\vec{F}_c, \\
  \label{eq:B}
  \pd{\vec{B}}{t} - \nabla\times (\vec{v} \times \vec{B}) &= 0,
  \end{align}
\end{subequations}
where $\rho$ is mass density, $\vec{v}$ is fluid velocity, $\vec{B}$ is magnetic field, $p_t = p + B^2/2$ is the total (gas + magnetic) pressure, $E = \rho\epsilon + \rho v^2/2 + B^2/2$ is the total energy density while $\tens{I}$ is the identity tensor. 

Thermal conduction effects, important for dilute gases and plasmas, are accounted for by the additional term on the right hand side of the total energy equation (\ref{eq:energy}). 
The thermal conduction flux $\vec{F}_c$ smoothly varies between classical and saturated regimes, 
\begin{equation}\label{eq:Fc}
  \vec{F}_{\rm c} = \frac{q}{q + F_{\rm class}} \vec{F}_{\rm class},
\end{equation}
where $F_{\rm class}$ and $q$ are, respectively, the magnitudes of the classical thermal conduction flux \citep{Cowie1977,Balsara2008} 
\begin{equation}\label{eq:Fclass}
\bm{F}_{\rm class} =  - \kappa_\parallel \hvec{b} ( \hvec{b} \cdot \nabla T)
\end{equation}
and the saturated flux
\begin{equation} \label{eq:Fsat}
  q  = 5\phi\rho c_{\rm iso}^{3}  \,.
\end{equation}
In Eq. (\ref{eq:Fclass}), $\kappa_\parallel$  is the thermal conductivity parallel 
to the direction of the magnetic field (a function of temperature; for expressions see pp. 38 of \citealt{NRLPF}) and $\hvec{b} \equiv \vec{B}/B$ is the magnetic field unit vector.
We do not consider thermal conduction perpendicular to magnetic field lines because it is typically much smaller, but including it is straightforward.
In absence of magnetic fields, the classical heat flux is simply given by (with $\kappa=\kappa_\parallel$)
\begin{equation}\label{eq:Fclass_HD}
  \vec{F}_{\rm class} = -\kappa\nabla T \,.
\end{equation}

Saturation of heat flux (Eq. \ref{eq:Fsat}) comes into play when the mean free path of the electrons is large compared to the temperature scale-height $L_T = T/|\nabla T|$ \citep{Cowie1977}.
The saturated heat flux is given by $q\hvec{F}_{\rm class}$ with $q$ specified in Eq. (\ref{eq:Fsat}), while $\hvec{F}_{\rm class} = \vec{F}_{\rm class}/F_{\rm class} 
= -\hvec{b}~{\rm sgn}( \hvec{b}\cdot \nabla T)$ is a unit vector along the local magnetic field line but down the temperature gradient.
In Eq. (\ref{eq:Fsat}) the parameter $\phi$ is an uncertainty factor of the order of unity (taken to be 0.3) while $c_{\rm iso}\equiv (p/\rho)^{1/2}$ is the isothermal sound speed. 
In the large temperature gradient limit, thermal conduction is described by a hyperbolic operator.
On the contrary, for $F_{\rm class}/q \to 0$, thermal conduction is described by a parabolic (diffusion) operator.
The flux given in Eq. (\ref{eq:Fc}) is a harmonic mean of the fluxes in the two regimes (weighted toward the smaller of the two) and reflects the mixed parabolic/hyperbolic mathematical nature of the underlying differential operators.
Eqs. 2.10 \& 2.10 in \citet{Balsara2008} gives some other ways of combining saturated and  classical heat fluxes.
For more details on our implementation of saturated conduction see Appendix A of \citet{Mignone2012}.

Note that thermal conduction described by Eq. (\ref{eq:Fc}) is very similar to the cosmic ray streaming equation (see Eqs. 1.1, 1.2 \& 6.1 in \citealt{Sharma2010}) in 
the sense that at temperature extrema, where gradient is zero and heat flux is not saturated, the behaviour is diffusive but at other points (in the long mean free 
path regime) it is hyperbolic. The harmonic mean used in Eq. (\ref{eq:Fc}) is analogous to the regularisations of the cosmic ray streaming equation at 
extrema proposed in \citet{Sharma2010}.

\section{Numerical Approaches for Anisotropic Thermal Conduction}
\label{sec:num}
%
%
In this section we review the various methods for implementing anisotropic thermal conduction.

We shall use the PLUTO code \citep{Mignone2007, Mignone2012} to solve Eqs. (\ref{eq:mass}-\ref{eq:B}) with the thermal conduction flux given by Eq. (\ref{eq:Fc}).
In PLUTO, STS schemes have been implemented in an operator split fashion so that we first advance Eqns. (\ref{eq:mass}-\ref{eq:B}) \textit{without} the conduction flux using standard Godunov-type methods, followed by the solution of
\begin{equation}\label{eq:scalar_diffusion}
  \pd{(\rho\epsilon)}{t} = - \nabla\cdot\vec{F}_c,
\end{equation}
by means of the STS/explicit approach.

A conservative approach is used during both steps so that the same flux is used for adjacent grid cells sharing a face.
During the parabolic step, Eq. (\ref{eq:scalar_diffusion}) is discretised using standard second-order finite differences:
\begin{equation}
  -\nabla\cdot\vec{F}_c \approx - \sum_d
  \frac{(\hvec{e}_d\cdot\vec{F}_c)_{i+\HALF} -
        (\hvec{e}_d\cdot\vec{F}_c)_{i-\HALF}}{\Delta h_d}
\end{equation}
where $d=x,y,z$ spans across dimensions, $\Delta h_d = (\Delta x,\,\Delta y,\,\Delta z)$ and $\hvec{e}_d$ are unit vectors in the three directions.
The interface flux is computed using central differences for the temperature gradient and an upwind scheme for the saturated heat term, as shown in appendix A of \cite{Mignone2012}.

\subsection{Explicit methods}
%

In the standard explicit approach (forward in time centred in space) the time step required to advanced the full system (\ref{eq:mass})-(\ref{eq:B}) is given by the shortest timescale:
\begin{equation}\label{eq:dt}
  \Delta t = \min\Big(\Delta t_{\rm h},\,\Delta t_{\rm p}\Big)\,,
\end{equation}
where
\begin{equation}\label{eq:dt_h}
  \Delta t_{\rm h}
  = \frac{C_h}
         {\DS \max\left(  \frac{\lambda_x}{\Delta x}
                        + \frac{\lambda_y}{\Delta y} 
                        + \frac{\lambda_z}{\Delta z}\right)}                         
\end{equation}
is the MHD/hydro (hyperbolic) time step whereas
\begin{equation}\label{eq:dt_exp}
  \Delta t_{\rm p}
  = \frac{C_p}
         {\DS \max\left(  \frac{\chi_x}{\Delta x^2}
                        + \frac{\chi_y}{\Delta y^2} 
                        + \frac{\chi_z}{\Delta z^2}\right)}                         
\end{equation}
is the diffusion time scale.
In Eqs. \ref{eq:dt_h} and \ref{eq:dt_exp}, the maximum is taken over the whole computational domain, $C_h \le 1$ is the hyperbolic Courant number, $C_p \le 1/2$ is the parabolic Courant number, $\lambda_{d} = |v_d| + c_{d,f}$, $c_{d,f}$ is the fast magneto-sonic speed in the $d$ direction, $\Delta x, \Delta y$ and $\Delta z$ are the mesh spacings in the three directions while $\chi_x,\chi_y$ and $\chi_z$ are the thermal diffusivities with the dimensions of $[{\rm length}]^2/[{\rm time}]$.
Adopting an ideal EoS ($p=[\gamma-1]\rho \epsilon$) one obtains, from Eq.  (\ref{eq:scalar_diffusion}) together with Eq. (\ref{eq:Fclass_HD}) or Eq. (\ref{eq:Fclass}):
\begin{equation}
  \chi_d = \left\{\begin{array}{ll}
  \DS \kappa\frac{\gamma-1}{\rho} & \qquad \mathrm{(HD)} \\ \noalign{\medskip}
  \DS \kappa_{\parallel}b_d\cos\theta \frac{\gamma-1}{\rho} & \qquad \mathrm{(MHD)}
  \end{array}\right.
\end{equation}
where $b_d = B_d/B$ is the component of the magnetic field unit vector in the direction $x_d$ and $\theta$ is the angle between $\vec{B}$ and $\nabla T$ while $d=x,y,z$
(note that, owing to operator splitting, density can be considered constant during the diffusion step).

Various explicit methods are discussed in detail in section 2 of \citet{sharma2007preserving}.
A notable symmetric explicit scheme with a very small numerical diffusion perpendicular to the local field lines is described in \citet{2005JCoPh.209..354G}.
This method has been used widely in modelling fusion plasmas, in which the temperature and field gradients are relatively gentle.
However, for highly nonlinear problems with large temperature gradients, this method gives unphysical temperature oscillations (see top right panel in Fig. 6 of \citealt{sharma2007preserving}).

It must emphasised that, for large grid resolutions and/or thermal conductivities, the diffusion timescale can become much shorter than the MHD/hydro time step (e.g., in galaxy clusters; \citealt{2014MNRAS.439.2822W,yang2015interplay}), that is $\Delta t_{\rm p} \ll \Delta t_{\rm h}$.
A fully explicit method can thus become very inefficient.
In these cases the ability to take larger steps than the stability limit (Eq. \ref{eq:dt}) is highly desirable.

A slight improvement over explicit time-stepping is obtained via sub-cycling, in which the MHD module is evolved with $\Delta t_{\rm h}$ but the conduction module is applied for a number of sub-steps ($N_{\rm sub} \approx \Delta t_{\rm h}/\Delta t_{\rm p}$; assuming $\Delta t_{\rm p} < \Delta t_{\rm h}$), each using the stable time step $\Delta t_{\rm p}$.While this approach is faster than the fully explicit approach, as we show in section \ref{sec:STS}, we can do much better than this using super-time-stepping.

\subsection{Implicit methods}
%
\label{sec:implmeth}
Implicit methods are very attractive because their time step is not limited by the stability limit (\cb{Eq.} \ref{eq:dt}).
A larger time step only leads to a slight loss in accuracy.
Also, for very high resolution simulations (i.e., with very small $\Delta x$), the CFL time step (Eq. \ref{eq:dt_h}) becomes much longer than the stability time step with diffusion (Eq. \ref{eq:dt_exp}).
For this reason implicit and semi-implicit schemes for diffusion are very popular (e.g., \citealt{2011JCoPh.230.4899S,2016A&A...585A.138D,kannan2015accurately,Pakmor:2016}).

The key problem with implicit methods is that they involve solving a sparse matrix equation, which is difficult to solve in parallel in a scalable 
way. Even a tridiagonal matrix equation which results from the semi-implicit method of \citet{2011JCoPh.230.4899S} is non-trivial to solve in parallel.
The MHD codes using implicit diffusion typically use sparse matrix libraries such as Hypre\footnote{\textit{http://acts.nersc.gov/hypre/}} or 
PETSc\footnote{\textit{http://www.mcs.anl.gov/petsc/}} to solve the matrix equation iteratively. 
This global solution approach is fundamentally different from the explicit update in which information only propagates across nearest grid 
cells in a single time step. Most importantly, even the scalable implicit methods for diffusion with MHD 
do not scale on more than $\sim$1000 processors on  massive distributed memory supercomputers (e.g., see Fig. 11 in \citealt{Caplan:2017}). 
In contrast, the STS methods that we advocate (in particular, RKL-STS) in this paper, being explicit, are trivial to parallelise 
and show good scaling up to tens of thousands of processors (see section \ref{sec:par_scaling}).

\subsection{Super-time-stepping: RKC, RKL, AAG}
\label{sec:STS}
%

Super-time-stepping (STS) is a way of choosing explicit time steps that are on-average much longer than the explicit stability time step 
(Eq. \ref{eq:dt_exp}).
The anisotropic diffusion equation has non-positive eigenvalues; i.e., none of the Fourier modes grow with time.
Following \citealt{meyer2014stabilized} \RefC{(see their section 2)}, we can write the diffusion equation in a semi-discrete ODE form
\begin{equation} \label{eq:ODE}
  \frac{du}{dt} = \tens{M} u,
\end{equation}
\RefC{
where, for a linear problem, $\tens{M}$ is a symmetric constant coefficient matrix which results from the discretization of the parabolic operator.
The eigenvalues $\lambda$ of such an operator are non-positive and real.}

The STS method can indeed be thought of as a multi-stage Runge-Kutta method in which the intermediate stages are chosen for stability, rather than for a higher order accuracy. 
The update of an $s-$stage scheme is quantified in terms of an amplification factor, $R_s$, defined as
\begin{equation}
u(t+\tau) = R_s(\tau \tens{M}) u(t).
\end{equation}

\RefC{
We now define a stability polynomial for an $s$-stage STS scheme, 
$$
R_{s}(\tau \lambda) \equiv \prod_{i=1}^s (1+\lambda \Delta t_i)
$$ 
with $\sum_{j=1}^s \Delta t_j = \tau$,
and we impose $|R_{s}(\tau \lambda)| \leq 1$  for all values of $\lambda$ between the largest negative eigenvalue of $\tens{M}$ and 0 in order to ensure stability.}
Temporal accuracy is achieved by matching  terms in the stability polynomial with the expansion of the analytic solution of Eq. (\ref{eq:ODE}),
\begin{equation}
\label{eq:tayexp}
u(t+\tau) = \left ( 1 + \tau \tens{M} + \frac{1}{2} (\tau \tens{M})^2 + ... \right) u(t).
\end{equation}
For first (second) order STS the stability polynomial only matches the above analytic expression \RefC{up to} the linear (quadratic) term. 

The sub-step sequence in STS was originally chosen by writing the stability polynomial, $R_s$ (a polynomial \RefC{of degree $s$} in $\lambda$), in form of a shifted Chebyshev polynomial  (CP; e.g., \citealt{verwer1990convergence}); we refer to this scheme as RKC (Runge-Kutta Chebyshev).
The sub-steps in a super-step are chosen to exploit the recursion properties of CPs.
Since the absolute value of CPs is always $\leq 1$ if their argument lies in $[-1,1]$, the magnitude of amplification factor after a super-time step is $\leq 1$.
Since CPs attain unity within $[-1,1]$, RKC-STS requires damping for the method to work for practical diffusion problems.

Recently, \citet{meyer2012second,meyer2014stabilized} have proposed Legendre polynomials (LPs) as the basis for the construction of robust STS schemes.
\RefC{
For a general $s$-stage RKL scheme, the stability polynomial is chosen to be of the form
\begin{equation}
\label{eq:stabpara}
R_s(\tau \lambda) = a_s + b_s \legP{s}{w_0 + w_1 (\tau \lambda)}
\end{equation}
where $w_0=1$ is chosen for all RKL schemes. The LPs, $\legP{j}{x}$ obey a three point recursion relationship given by
\begin{equation}
\label{eq:recLegP}
 \legP{j}{x} = \left(\frac{2j - 1}{j}\right) x \legP{j-1}{x} - \left(\frac{j-1}{j}\right)\legP{j-2}{x}.
\end{equation}
RKL-2 method enforces the stability polynomial at sub-stage $j$ to be $a_j+b_j \legP{j}{\tau \tens{M}}$
and uses the property of LPs that their absolute value is bounded by unity if their argument lies in (-1,1).
Following \citet{meyer2014stabilized}, in Appendix \ref{sec:exprkl2} we 
illustrate the second order temporal accuracy of the RKL-2 scheme along with explicit formulae for 
an $s = 3$ stage scheme. 
}

The key advantage of RKL schemes over RKC is that they are more robust because LPs, unlike CPs, are always smaller than unity in magnitude if their argument lies in (-1,1).	
Thus, there is no need of an explicit damping parameter ($\nu$ in Eqs. 9 \& 10 of \citealt{alexiades1996super}). Moreover, \citet{meyer2014stabilized} show 
that RKL has superior linear stability and monotonicity properties compared to RKC. 

\RefC{For stability, the argument of the LP in Eq. (\ref{eq:stabpara}) should be $\geq -1$ (note that for a parabolic operator the eigenvalues are
 non-positive); i.e., $1 - w_1 \tau |\lambda|_{\rm max} \geq -1$ or $\tau \leq 2/(|\lambda|_{\rm max} w_1)$, where $|\lambda|_{\rm max}$ is the 
 eigenvalue of $\tens{M}$ with largest absolute value. Note that $w_1=2/(s^2+s)$ and $4/(s^2+s-2)$ for RKL-1 and RKL-2, respectively.}
Given a super-time-step that can be conveniently chosen as the hyperbolic CFL time step $\Delta t_{\rm h}$ (Eq. \ref{eq:dt_h}),  the number of 
sub-stages in RKL-STS is given by (see Eq. 20, 21 in 
\citealt{meyer2014stabilized})
\begin{equation} \label{eq:RKL-s}
  \begin{array}{lcll}
    \Delta t_{\rm h} &=&\DS \Delta t_{\rm p} \frac{(s^2+s)}{2}  & \text{for RKL-1,} 
    \\ \noalign{\medskip}
    \Delta t_{\rm h} &=&\DS \Delta t_{\rm p} \frac{(s^2+s-2)}{4}& \text{for RKL-2}.
  \end{array}
\end{equation}
where, abbreviations RKL-1 and RKL-2 stand for the first and second order Runge-Kutta-Legendre schemes.\RefC{We assume here that $s>1$; 
$s=1$ corresponds to a standard forward Euler update.} 
In this paper, RKL without a qualifier refers
to RKL-2. We explicitly mention if we are using the first order method. It is possible to make the scheme
even faster by taking multiple $\Delta t_{\rm h}$ as the size of a super-step but of course one needs to worry about accuracy.

Likewise, for AAG-STS, the largest number of first order Euler sub-steps $s$ satisfies  (see Eq. 10 in \citealt{alexiades1996super})
\begin{equation}\label{eq:AAG-STS}
  \Delta t_{\rm h} = \Delta t_{\rm p} \frac{s}{2\sqrt{\nu}}
                     \left[\frac{(1+\sqrt{\nu})^{2s} - (1-\sqrt{\nu})^{2s}}
                                {(1+\sqrt{\nu})^{2s} + (1-\sqrt{\nu})^{2s}}\right]
   .
\end{equation}

The maximum $s$ for STS schemes gives a speed-up of the conduction module scaling as $(\Delta t_{\rm h}/\Delta t_{\rm p})^{1/2}$.
The speed-up is marginally better for RKC (compare Eqs. 18, 19 in \citealt{meyer2012second}) although RKL has better stability 
(as shown in section \ref{sec:tests}), and hence is much more attractive.

It is instructive to compute the effective parabolic CFL number $C_p=\chi \Delta t/\Delta x^2=\Delta t_{\rm h}/2 \Delta t_{\rm p}$ as a function of the number of sub-steps $s$.
In the limit of large $s$ ($s\gg 1$), the RKL schemes (\ref{eq:RKL-s}) yield
\begin{equation}
  C_p \to \frac{s^2}{4} \quad\mathrm{(RKL-1)} \,,\qquad
  C_p \to \frac{s^2}{8} \quad\mathrm{(RKL-2)}.
\end{equation}
However, in the same limit, AAG-STS yields an effective gain which depends on $\nu$:
\begin{equation}\label{eq:STS-limit}
  C_p \to \left\{\begin{array}{lcl}
        \DS \frac{s^2}{2} & \quad \mathrm{for} & s \lesssim 1/(2\sqrt{\nu}).
        \\ \noalign{\medskip}
        \DS \frac{s}{4\sqrt{\nu}} & \quad \mathrm{for} & s \gtrsim 1/(2\sqrt{\nu}).
        \end{array}\right.
\end{equation}
This equation sets an upper limit on the gain that can be reached using the traditional AAG-STS scheme: no substantial gain is obtained when $s\gtrsim 1/(2\sqrt{\nu})$.

\begin{table}
\caption{Expected Flops relative to MHD evolution}
\resizebox{0.48 \textwidth}{!}{%
\begin{tabular}{c c c c c }
\hline
MHD & explicit & sub-cycling & STS & implicit$^\dag$ \\
\hline
1 & $(1+r^{-1})t^{-1}$ &  $1+t^{-1}r^{-1}$ & $1+t^{-1/2}r^{-1}$ & $1+r^{-1}$ \\
\hline
\end{tabular}}
\label{tab:scaling}
\textbf{Notes:} $r$ is the ratio of number of Flops needed for the MHD module and the conduction module; $t=\Delta t_{\rm p}/\Delta t_{\rm h}$ 
(taken to be $<1$; Eqs. \ref{eq:dt_exp} \& \ref{eq:dt_h}). \\
$^\dag$ here it is assumed that a single implicit update for conduction takes the same number of Flops as a single explicit update; the implicit update is 
likely to be more expensive and the number of expected Flops is larger.
\end{table}

As mentioned earlier, the key advantage of STS methods is their simplicity, parallel scaling on massively distributed clusters, and their large
speed-up compared to the explicit method. Table \ref{tab:scaling} shows the expected scaling of Flops (equivalently, run-time) using various 
conduction schemes relative to pure MHD evolution.
Here, $t=\Delta t_{\rm p}/\Delta t_{\rm h}$ (assumed to be $<1$) and $r$ is the ratio of Flops (floating point operations) for MHD and conduction 
modules. The typical value of $r$ is a few (say 5; this is relatively independent of the physical set-up and dimensionality) but $t$ can be smaller 
than 0.01 (this depends on the physical set-up). For these typical values the ratios of Flops required in Table \ref{tab:scaling} (for MHD, explicit, sub-cycling, STS, implicit) 
become: 1, 120, 21, 3, 1.2.
Therefore, STS promises a good speed-up compared to the explicit method, and is competitive relative to implicit methods (especially 
given the better parallel scaling of the former).

\subsection{Monotonicity in Multi-Dimensions}
%

A serious problem for anisotropic 
diffusion, in presence of large gradients, is that the temperature can behave non-monotonically. The use 
of limiters for interpolating transverse temperature gradients has proved useful in preventing negative 
temperatures in such cases (\citealt{sharma2007preserving}; see Appendix \ref{app:tr_lim} for details).

The use of limiters to calculate the transverse temperature gradients (Eq. \ref{eq:qT}) has shown to improve the robustness of all 
schemes (e.g., see \citealt{sharma2007preserving,2011JCoPh.230.4899S}). However, we show in section \ref{sec:ring_diffusion} that limiters reduce the accuracy of STS schemes for a large number of sub-steps. Given this, we keep limiters as an option to be used only when temperature gradients are large and the wrong sign of the heat flux gives a negative temperature in some grid cells.

\section{Numerical Tests: effectiveness of RKL-STS}
\label{sec:tests}
%
%
%

In the following we present selected numerical benchmarks.
We adopt an ideal equation of state so that
\begin{equation}
  \rho\epsilon = \frac{p}{\gamma - 1}
\end{equation}
where $\gamma$ is the ratio of specific heats which equals $5/3$ unless otherwise stated.

\subsection{Scalar Diffusion Equation}
\label{sec:scalar_tests}
%
%

In this section we consider numerical tests based on the solution of the scalar diffusion equation
\begin{equation}\label{eq:scalar_diffusion2}
  \pd{(\rho\epsilon)}{t} = - \nabla\cdot\vec{F}_c,
\end{equation}
where $\vec{F}_c$ is given by Eq (\ref{eq:Fc}).
Other fluid variables are not evolved in time.

In all computations, we assign the super-step $\Delta t_{\rm h}$ and compute the number of sub-steps by solving either Eq. \ref{eq:RKL-s} (for RKL) or Eq. \ref{eq:AAG-STS} 
(for AAG-STS) for $s$ and then rounding the solution to the next larger integer, i.e., $s\,\to\, 1 + {\rm floor}(s)$.

\subsubsection{Gaussian diffusion in 1-D}
\label{sec:gaussian1d}
%
%
%

A standard one dimensional diffusion test is useful to compare the stability and accuracy of different methods described in section \ref{sec:STS}.
For this purpose, we solve the standard heat equation (without saturation) for temperature:
\begin{equation}\label{eq:generic_diff}
  \frac{\partial}{\partial t} T(x,t) = \frac{\partial}{\partial x} \left ( \kappa \frac{\partial T(x,t)}{\partial x} \right ),
\end{equation}
using a constant diffusion coefficient $\kappa = 1$.
In the PLUTO code, this is equivalent to solving Eq. \ref{eq:scalar_diffusion} by setting $\rho = 1$ and $\gamma=2$, so that $\rho\epsilon=p=T$ in code units.

The initial condition consists of a Gaussian temperature profile $T(x,0) = \exp(-x^2/2\sigma^2)$ inside the domain $x\in[-2,2]$.
Then Eq. \ref{eq:generic_diff} admits the well-known analytical solution
\begin{equation}\label{eq:gaussian_exact}
  T^{\rm ref}(x, t) = \frac{1}{\sqrt{1 +
      \frac{2\kappa t}{\sigma^2}}}\exp\left[-\frac{x^{2}}{2\sigma^{2}\left(
         1 + \frac{2\kappa t}{\sigma^{2}}\right)}\right] \,.
\end{equation}
Boundary conditions are specified using the exact solution.

\begin{figure}
\begin{center}
\includegraphics[width=1\columnwidth]{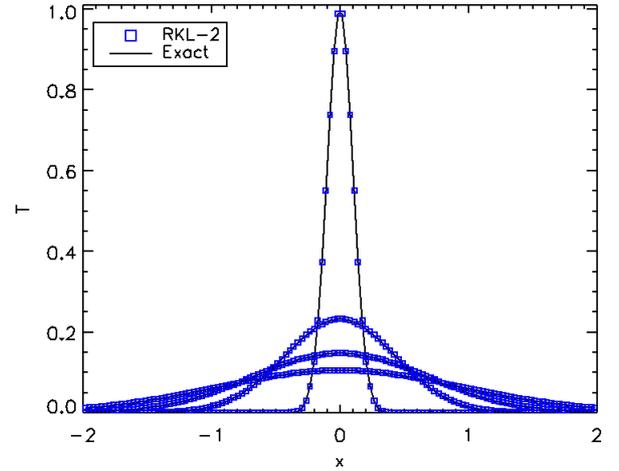}
\caption{Temperature profiles (blue square symbols) for the 1D Gaussian diffusion
         problem at $t = 0,\, 0.08,\, 0.22,\, 0.45$ obtained with RKL-2 using
        $N_x=128$ grid zones and a parabolic CFL number $C_p=10$.
         The exact analytical solution (Eq. \ref{eq:gaussian_exact}) is given
         by the black solid line.}
\label{fig:gauss_diff}
\end{center}
\end{figure}

Figure \ref{fig:gauss_diff} shows the computed temperature profile (symbols) versus the analytical solution (Eq. \ref{eq:gaussian_exact}; solid line) 
at different times using RKL-2 on $N_x=128$ grid points and a parabolic CFL number $C_p~(\equiv \Delta t_{\rm h}/2 \Delta t_{\rm p}) = 10$.
As expected, the initial peak of the \textit{Gaussian} temperature profile spreads symmetrically about $x=0$ with time.

\begin{figure*}
\begin{center}
\includegraphics[width=0.325\textwidth]{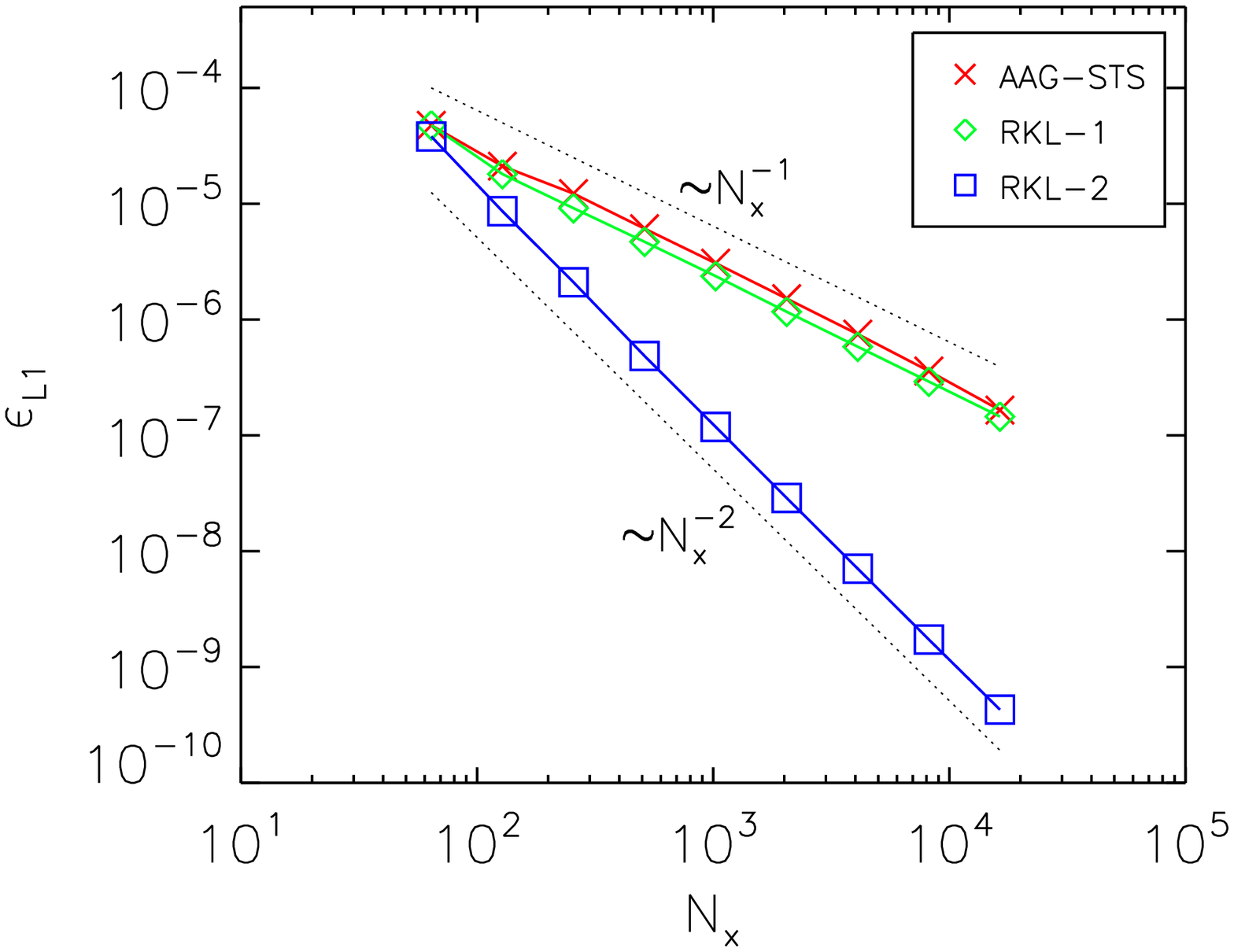}
\includegraphics[width=0.325\textwidth]{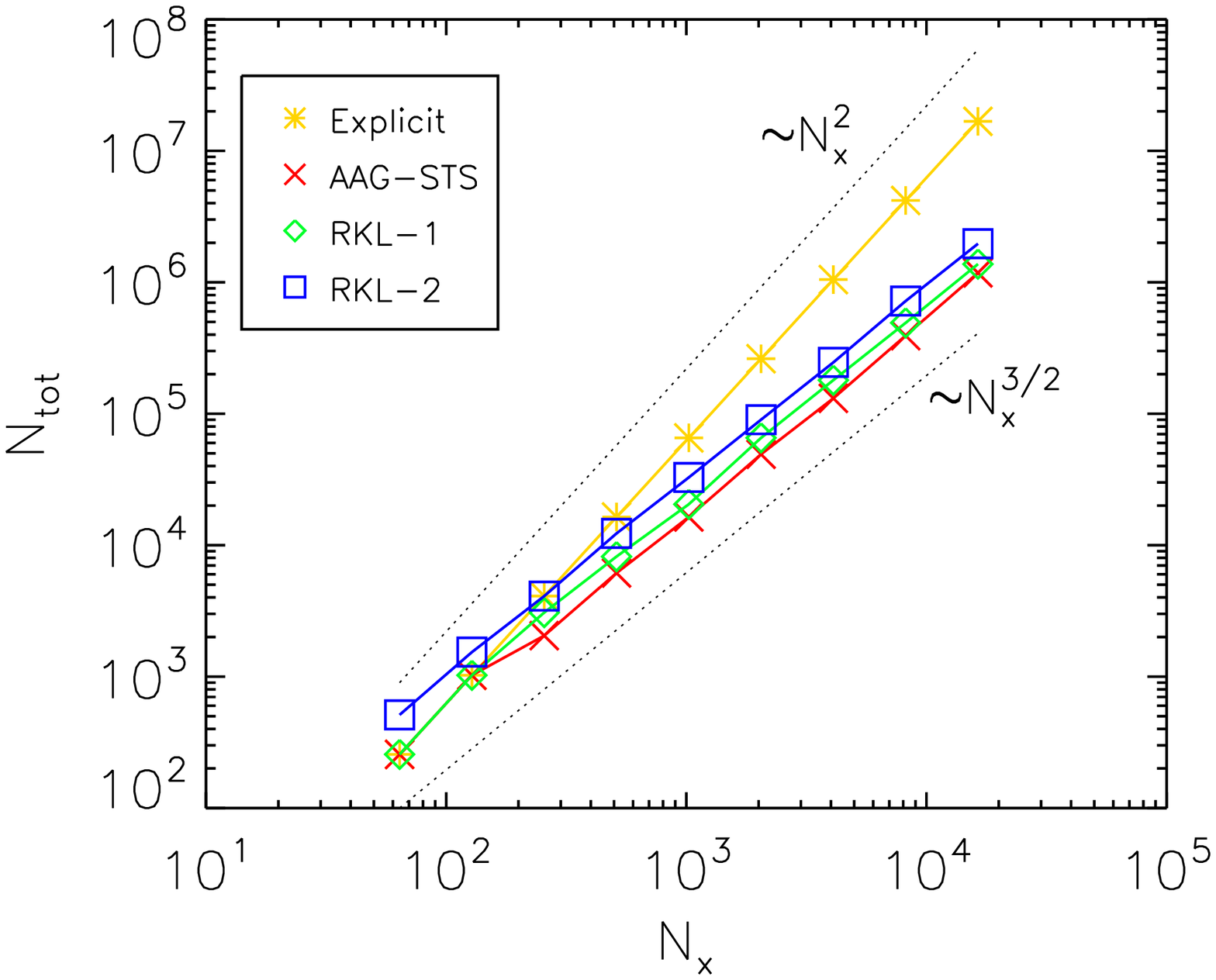}
\includegraphics[width=0.325\textwidth]{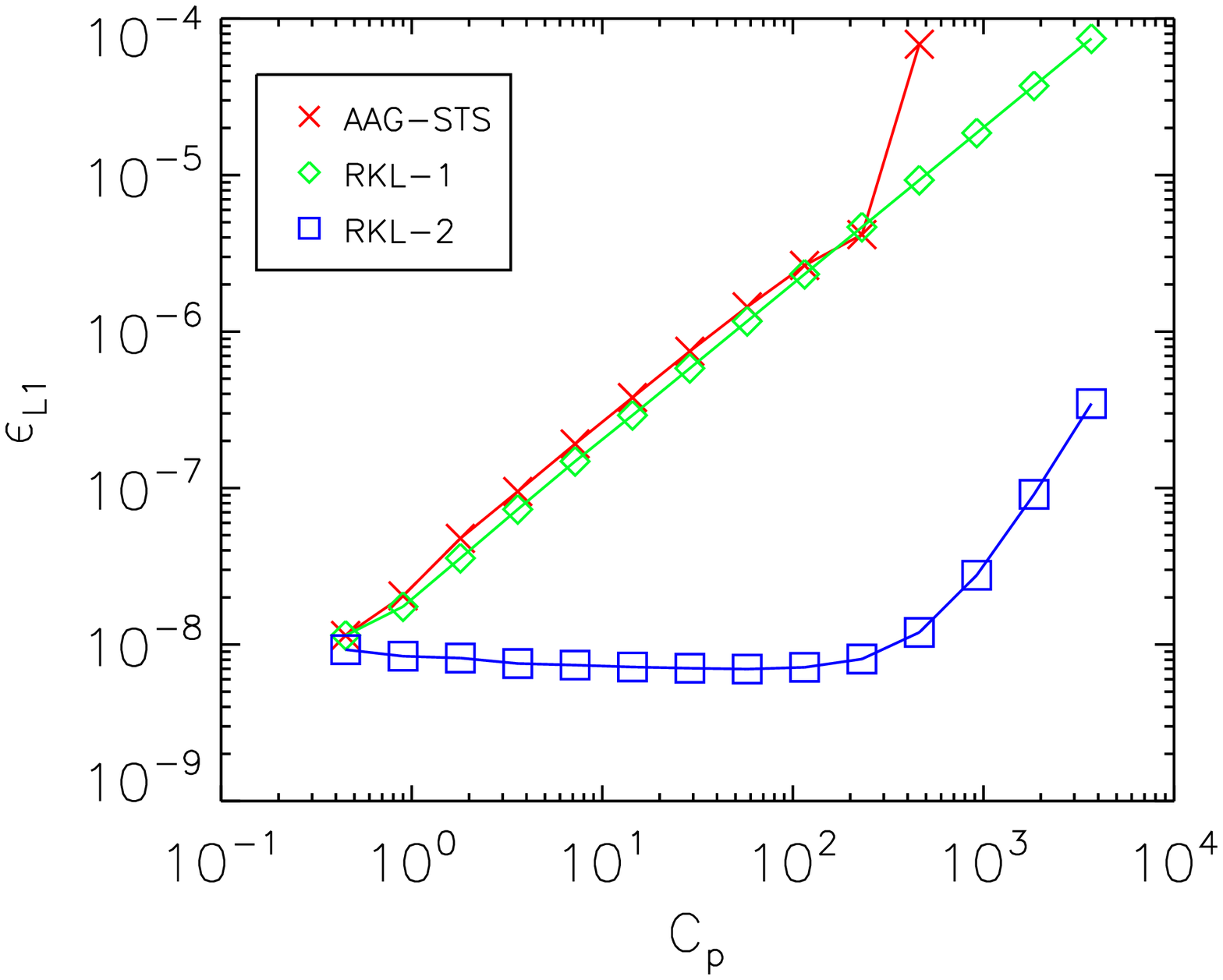}
\caption{\RefC{Error analysis for the 1D Gaussian test problem using
          different explicit schemes.}
	\textit{Left panel}: errors in L1 norm as functions of the grid resolution
         $N_x$ for AAG-STS (red), RKL-1 (green) and RKL-2 (blue) methods.
         The two dotted lines give the ideal scalings for first and
         second order schemes.
         \textit{Middle panel}: total number of steps (integration steps $\times$
         number of sub-steps) for the explicit 1st order scheme (yellow) and 
         AAG-STS, RKL-1 and RKL-2 (red, green and blue respectively).
         The computational workload follows the expected scaling of $N_x^2$
         (explicit) and $N_x^{3/2}$ (STS) shown by dotted lines.
         \textit{Right panel}: error in L1 norm as a function of the parabolic CFL 
         number $C_p = \kappa\Delta t_{\rm h}/\Delta x^2$ at the fixed resolution 
         $N_x=4096$. }
\label{fig:gauss_diff_err1}
\end{center}
\end{figure*}

In order to verify the order of accuracy of the various methods considered here, we first perform a resolution study by doubling the grid resolution from $N_{x}=64$ up to $N_x=16384$. 
We quantify the accuracy for different integration methods, AAG-STS (we use $\nu=10^{-3}$ unless otherwise stated), RKL-1, RKL-2, by estimating the error in $L1$ norm as
\begin{equation}\label{eq:gauss_emax}
  \epsilon_{L1} = \frac{1}{N_x}\sum_i \left|T^n_i - T^{\rm ref}(x_i,t^n)\right|.
\end{equation}
Since we use standard centred differencing for the second derivative in Eq. \ref{eq:generic_diff} with local error $O(\Delta x^2)$, we must ensure that 
the error in our simulation is not dominated by spatial discretisation.
To this end, we adjust the time step such that the ratio of  the time step and the grid spacing  is fixed; i.e.,
\begin{equation}\label{eq:gauss_constraint}
  \frac{\Delta t_{\rm h}}{\Delta x} = \frac{C_p \Delta x}{\kappa} = \frac{C_{p0}}{N_{x0}\kappa} = {\rm const},
\end{equation}
where $N_{x0} = 64$ and $C_{p0} = 0.45$ are, respectively, the number of zones and the parabolic CFL number at the lowest resolution.
This yields a local truncation error (for $T$) of $O(\Delta t^m) + O(\Delta t^3)$ where $m=2,3$ for a first or second order scheme, respectively.
Notice also that the parabolic CFL number increases linearly with the spatial resolution.

The left panel in Fig. \ref{fig:gauss_diff_err1} shows the errors obtained for AAG-STS, RKL-1 and RKL-2 methods together with the expected first and second-order convergence rates. 
From the figure one can verify that the order of accuracy for AAG-STS and RKL-1 is that of a $1^{\rm st}$ order method whereas RKL-2 converges as a $2^{\rm nd}$ order method, as expected.
We remark that integration with AAG-STS shows the occurrence of \textit{negative} temperatures during sub-steps for $N_x \ge 8192$ (corresponding to $s \gtrsim 30$), 
although the solution remains positive at the end of the super-step.
This can pose serious difficulties in more complex applications in which the diffusion coefficient is a nonlinear function of the temperature (e.g., Spitzer conductivity $\kappa\propto T^{5/2}$) requiring $T$ to be non-negative at all times.
On the contrary, RKL-1 and RKL-2 never exhibit such a behaviour (i.e., the solution remains \textit{positive} for all sub-steps; this property is enforced by construction in the RKL scheme as described at the end of section 2.2 of \citealt{meyer2014stabilized}).

In the middle panel of Figure \ref{fig:gauss_diff_err1} we plot the computational time (measured as the total number of steps and sub-steps) as a function of the number of grid points $N_x$.
For an explicit scheme, let $N_t \propto N_x^2$ be the total number of steps required to reach some final time step.
Then, from Eqs. \ref{eq:RKL-s} \& \ref{eq:STS-limit}, we expect the number of sub-steps using STS schemes to scale roughly as $s\propto \sqrt{\Delta t_{\rm h}/\Delta t_{\rm p}} \propto \sqrt{N_x}$.
The total number of steps (integration steps $\times$ number of sub-steps) is therefore expected to be $N_t\propto N_xs \propto N_x\sqrt{N_x}$.
This behaviour is verified in the middle panel of Figure \ref{fig:gauss_diff_err1} from which we conclude that STS techniques provide an effective \textit{asymptotic gain} over standard explicit time-stepping proportional to the \textit{square root} of the number of grid points.

Next, in the right panel of \RefC{Figure \ref{fig:gauss_diff_err1}}, we plot the L1-norm errors by changing the parabolic CFL number $C_p = \kappa\Delta t_{\rm h}/\Delta x^2$ for 
a fixed grid resolution $N_x = 4096$.
Although comparable for $C_p \lesssim 1$, the errors grow linearly with the CFL number for the first order schemes like AAG-STS and RKL-1. 
However, for $C_p \gtrsim 200$, AAG-STS becomes unstable and integration is no longer possible, while RKL-1 remains stable without showing any significant undershoot even during each cycle sub-step.
The second order scheme maintains approximately the same accuracy for $C_p \lesssim 10^3$ and the error starts to increase more rapidly for larger CFL although the solution remains well-behaved and positive at all times.

\begin{figure}
\begin{center}
\includegraphics[width=1\columnwidth]{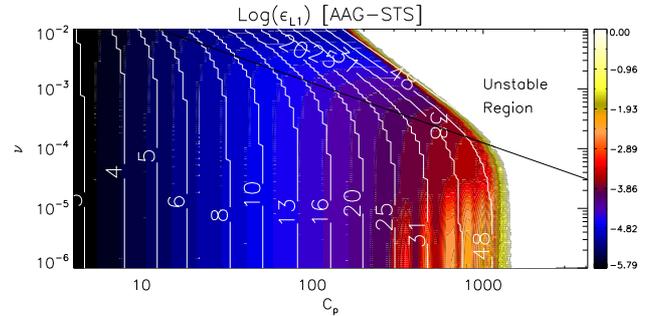}
\caption{Two-dimensional contour plot of the logarithm of the L1 error (Eq. \ref{eq:gauss_emax})
         \RefC{for the 1D Gaussian test problem} as a function
         of the parabolic CFL number $C_p$ and the $\nu$ parameter for the
         AAG-STS scheme with fixed resolution $N_x=1024$.
         Blue colours indicate smaller errors while red, orange and yellow denote
         increasingly larger errors and unstable behaviour.
         The number of sub-steps $s$ is over-plotted using white contour lines. 
         The value of $s$ depends on the CFL number which varies from 
         $4$ to $4096$.
         The thick solid line represents $\nu=1/(8C_p)$, the efficiency limit
         above which the parabolic CFL number grows linearly with the number
         of sub-steps (Eq. \ref{eq:STS-limit}).
         Notice that all computations fail when $s \gtrsim 50$.}
\label{fig:gauss_diff_err2}
\end{center}
\end{figure}

Finally we investigate the stability properties of the AAG-STS method alone, by varying both the parabolic 
CFL number and the $\nu$ parameter at the fixed grid resolution of $N_x=1024$ zones.
Fig. \ref{fig:gauss_diff_err2} shows the logarithm of the error as a function of $C_p$ and $\nu$.
The corresponding number of sub-steps $s$ is also over-plotted using white contour lines.
For sufficiently small values of the damping parameter ($\nu \lesssim 10^{-4}$), we see that the error rapidly increases when $C_p$ exceeds $\sim 200$ ($s\sim 20$) and computations eventually become unstable for $C_p\gtrsim 10^{3}$ ($s \gtrsim 45$), irrespective of the value of $\nu$.
For $s\lesssim 1/(2\sqrt{\nu})$, Eq. \ref{eq:STS-limit} shows that $C_p\approx s^2/2$ is independent of $\nu$.
For larger values of the damping parameter ($\nu \gtrsim 10^{-4}$), however, the number of steps required to complete the calculation at a given CFL number increases and therefore we observe a loss of efficiency that reduces the limiting CFL number from $C_p\approx 10^3$ (at $\nu\approx 10^{-4}$) to $C_p\approx 200$ (for $\nu \approx 10^{-2}$). 
Of course, increasing the number of sub-steps at a given CFL number leads to larger stability at the cost of extra computational work.
Finally, we point out that our temperature profiles are marginally affected by the numerical resolution, the effect of which is that of triggering (in case of 
numerical instability for $s \gtrsim$ 40) the growth of Nyquist ($k=\pi/\Delta x$) mode.

\subsubsection{Saw-tooth profile with saturated heat flux}
%

In the next example we compare the performance of the selected integration schemes by also including 
saturation of the heat flux.
More specifically, we consider the initial sawtooth temperature profile
\begin{equation}
  T(x,0) = 10 + 20\left[\frac{x}{10}
                         - \mathrm{floor}\left(\frac{x}{10} + \frac{1}{2}\right)\right]
\end{equation}
and solve Eq. \ref{eq:scalar_diffusion} by setting $\rho = 1$ and $\gamma = 2$ so that $\rho\epsilon=p=T$ in code units, as in the previous example.
The thermal conduction flux is given by Eq. \ref{eq:Fc} with $\vec{F}_{\rm class}$ and $q$ defined in Eq. \ref{eq:Fclass_HD} and Eq. \ref{eq:Fsat}, respectively, with $\kappa = 40$, $\phi = 0.3$.
Computations are carried out on the one-dimensional domain $x\in[-L_x/2,L_x/2]$ using $N_x = 400$ zones and periodic boundary conditions.
The total number of steps to reach some final time $t_{\rm stop}$ is therefore $N_{\rm step} = t_{\rm stop}/\Delta t_{\rm h}$, with $\Delta t_{\rm h}$ computed from the parabolic CFL number
\begin{equation}
  N_{\rm step} = \frac{t_{\rm stop}\kappa}{C_p}
  \left(\frac{N_x}{L_x}\right)^2
               = \frac{1600}{C_p} \,,
\end{equation}
where we have used $t_{\rm stop} = 0.1$.
\begin{figure}
\includegraphics*[width=0.5\textwidth]{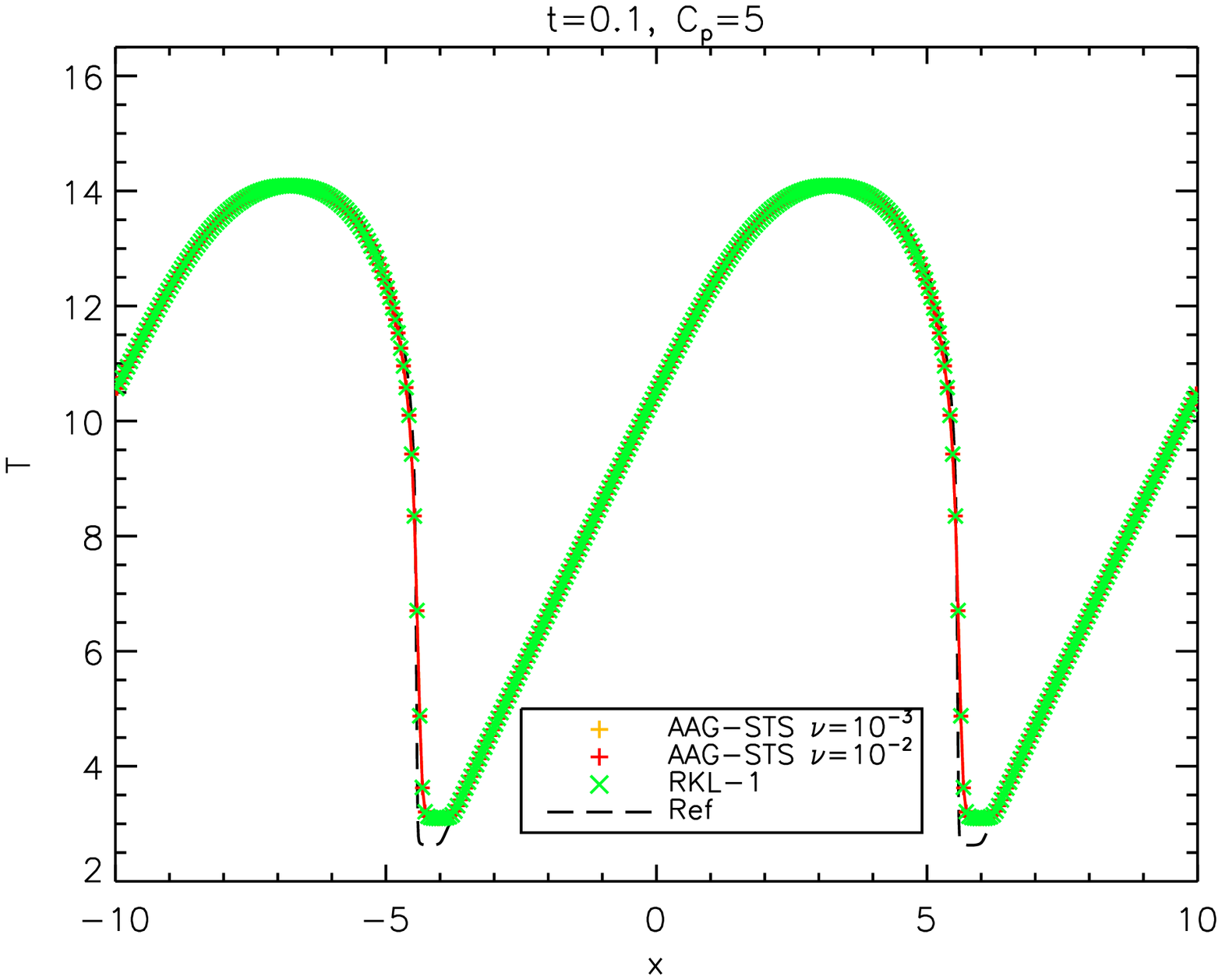}
\includegraphics*[width=0.5\textwidth]{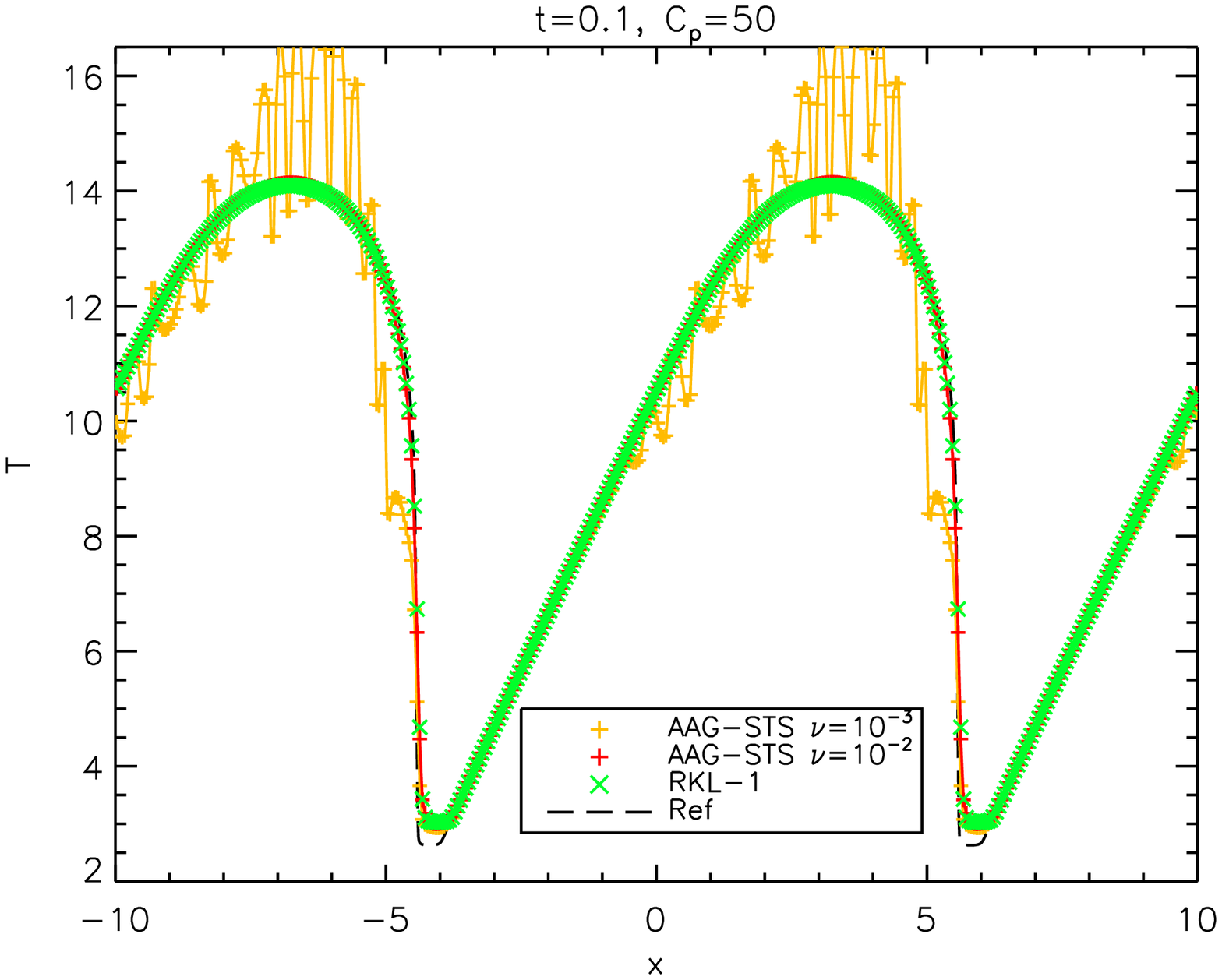}
\caption{Numerical solution of the sawtooth temperature profile with heat
         saturation at $t=0.1$.
         \textit{Top panel}: computations obtained with parabolic CFL number
          $C_p = 2$ for AAG-STS ($\nu=10^{-3},\, 10^{-2}$ orange and red,
          respectively), RKL-1 (green) and reference solution (black dashed line).
         \textit{Bottom panel}: same as before but using a parabolic CFL number
          $C_p = 50$. Spurious oscillations appear using AAG-STS with $\nu=10^{-3}$.}
\label{fig:sawtooth_comp}
\end{figure}
We perform two sets of computations corresponding to $C_p=5$ and $C_p=50$ and compare the results with a reference solution obtained on a much finer grid.
Results are shown in Fig. \ref{fig:sawtooth_comp}.
In the case of AAG-STS we employ $\nu = 10^{-3}$ (orange) and $\nu = 10^{-2}$ (red).

Away from extrema, the evolution is initially dominated by the contribution of the saturated flux only (since $\partial^2T/\partial x^2 = 0$ everywhere except where the first derivative is discontinuous). The diffusive part of the flux acts in those region where a change in slope is present.
For $C_p=5$, all methods yield well-behaved solutions with comparable errors whereas for $C_p=50$, high frequency spurious oscillations appear when using AAG-STS with a lower values of the damping parameter ($\nu = 10^{-3}$) as shown by the orange line in Figure \ref{fig:sawtooth_comp}.
Oscillations originate in proximity of the maxima where the second derivative of $T$ does not vanish and propagate downstream as the system evolves.
Increasing the grid resolution or the parabolic CFL number tends to amplify this unstable behaviour.
We note that oscillations disappear if saturation is not included.

On the other hand, computations remain stable when using RKL-1 (green) or RKL-2 (not shown) even for larger parabolic CFL numbers ($C_p \lesssim 200$).
For $C_p\gtrsim 200$ we observe larger numerical errors in the solution even with RKL. The production of temperature oscillations at maxima is analogous 
to numerical oscillations produced at extrema when numerically solving the cosmic ray streaming equation (see Fig. 2.1 in \citealt{Sharma2010}). Unlike in cosmic 
ray streaming, oscillations are not produced at temperature minima because of a smaller streaming speed ($\propto \sqrt{T}$). We have verified this dependence on streaming speed by running with a higher initial temperature (T = 30, instead of T = 10) and noticing oscillatory behavior both at temperature maxima and minima in the case of AAG-STS with $\nu = 10^{-3}$, while other cases show stable solutions.

\subsubsection{Ring diffusion}
\label{sec:ring_diffusion}
%
%

The 2-D Cartesian ring diffusion test problem, presented in \citet{2005ApJ...633..334P,sharma2007preserving}, is useful to study the monotonicity properties of 
various numerical schemes for anisotropic diffusion in presence of temperature discontinuities. Temperature discontinuities are fairly common in 
astrophysical fluids and plasmas, and an ideal numerical scheme should not lead to negative temperatures in presence of large temperature gradients.
For an explicit update with $\Delta t \leq \Delta t_{\rm p}$, \citet{sharma2007preserving} showed that temperature monotonicity is maintained 
if limiters are used to interpolate traverse 
temperature gradients (see Appendix \ref{app:tr_lim} for details); using arithmetic mean for interpolation leads to non-monotonicity in general. The use 
of limiters with an implicit/semi-implicit update does not strictly maintain monotonicity but improves monotonicity substantially compared with arithmetic 
averaging (see Figs. 4 \& 5 in \citealt{2011JCoPh.230.4899S}).

We numerically solve the anisotropic thermal diffusion equation, Eq. \ref{eq:scalar_diffusion} with $\kappa_\parallel=1$; saturation of heat flux is 
ignored for this test problem.
We set $\rho=1$, $\gamma=2$, and $\rho\epsilon=p=T$ in code units. The magnetic field lines are circular with $B_x=y/(x^2+y^2)^{1/2}$ and 
$B_y=-x/(x^2+y^2)^{1/2}$, and the fluid is static.
All variables except $T$ ($\rho$, $\bm{B}$, $\bm{v}$) are held fixed; $T$ evolves only because of anisotropic diffusion. The computational 
domain is $[-1,1]\times[-1,1]$, equal number of grid points ($N$) are used in the two directions, and periodic boundary condition is imposed on $T$. 
The initial condition on $T$ is
\begin{equation}\label{eq:T-ring}
  T = \left\{\begin{array}{ll}
        \DS 12 &  \mathrm{for}\quad\DS \frac{11\pi}{12} \leq \theta \leq
                                    \frac{13\pi}{12}, \quad 0.5 \leq r \leq 0.7,
        \\ \noalign{\medskip}
        \DS 10  &  \mathrm{otherwise},
        \end{array}\right.
\end{equation}
where $r=(x^2+y^2)^{1/2}$ and $\theta$ ($\tan \theta=y/x$) are polar coordinates. The magnetic field lines are not aligned with the Cartesian grid and the transverse heat 
flux (Eq. \ref{eq:qT}) is non-zero. The anisotropic diffusion equation is evolved \RefC{until} $t=1$ using 
different schemes.

\begin{figure*}
  \includegraphics[scale=0.42]{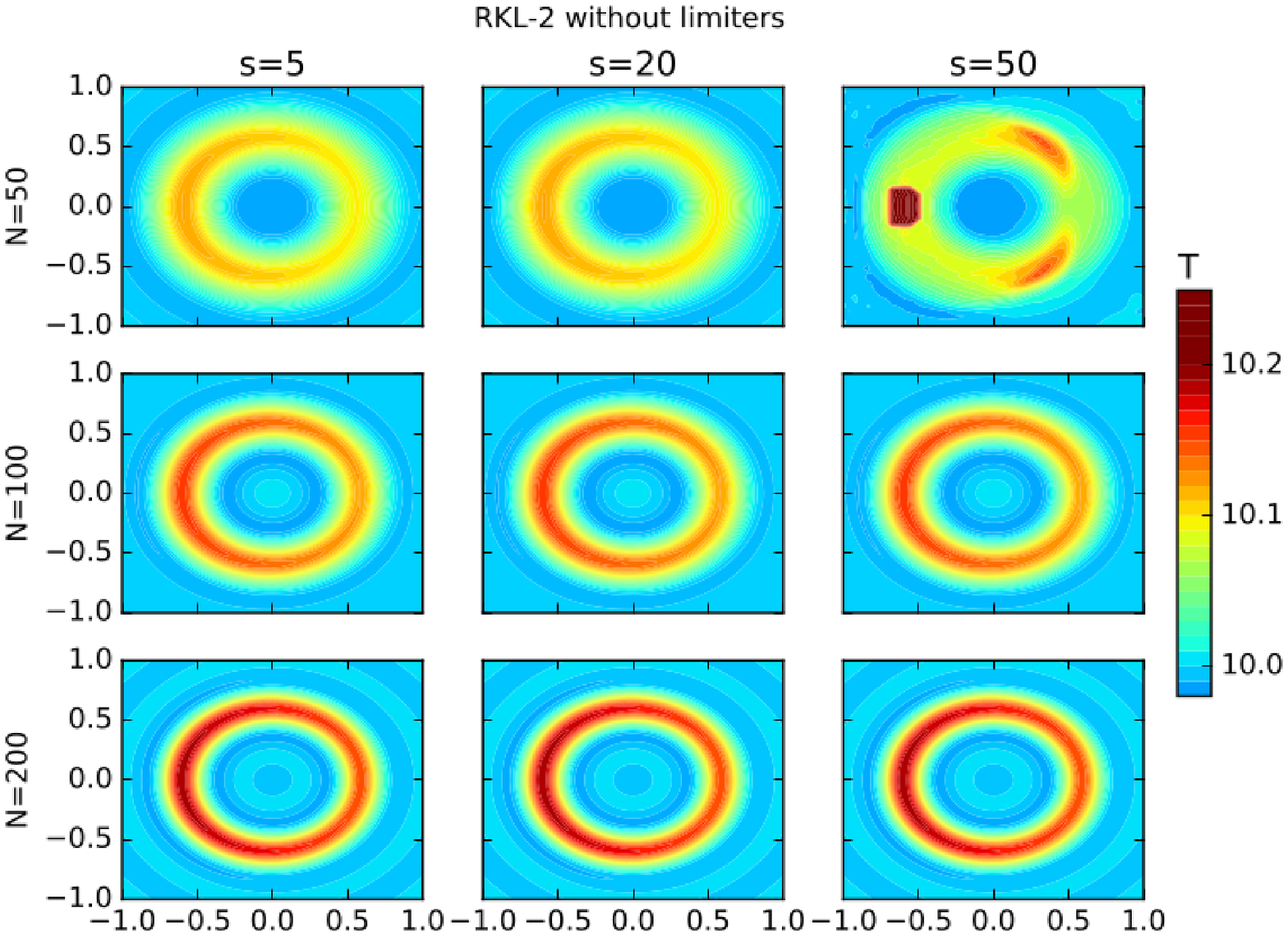}
  \includegraphics[scale=0.42]{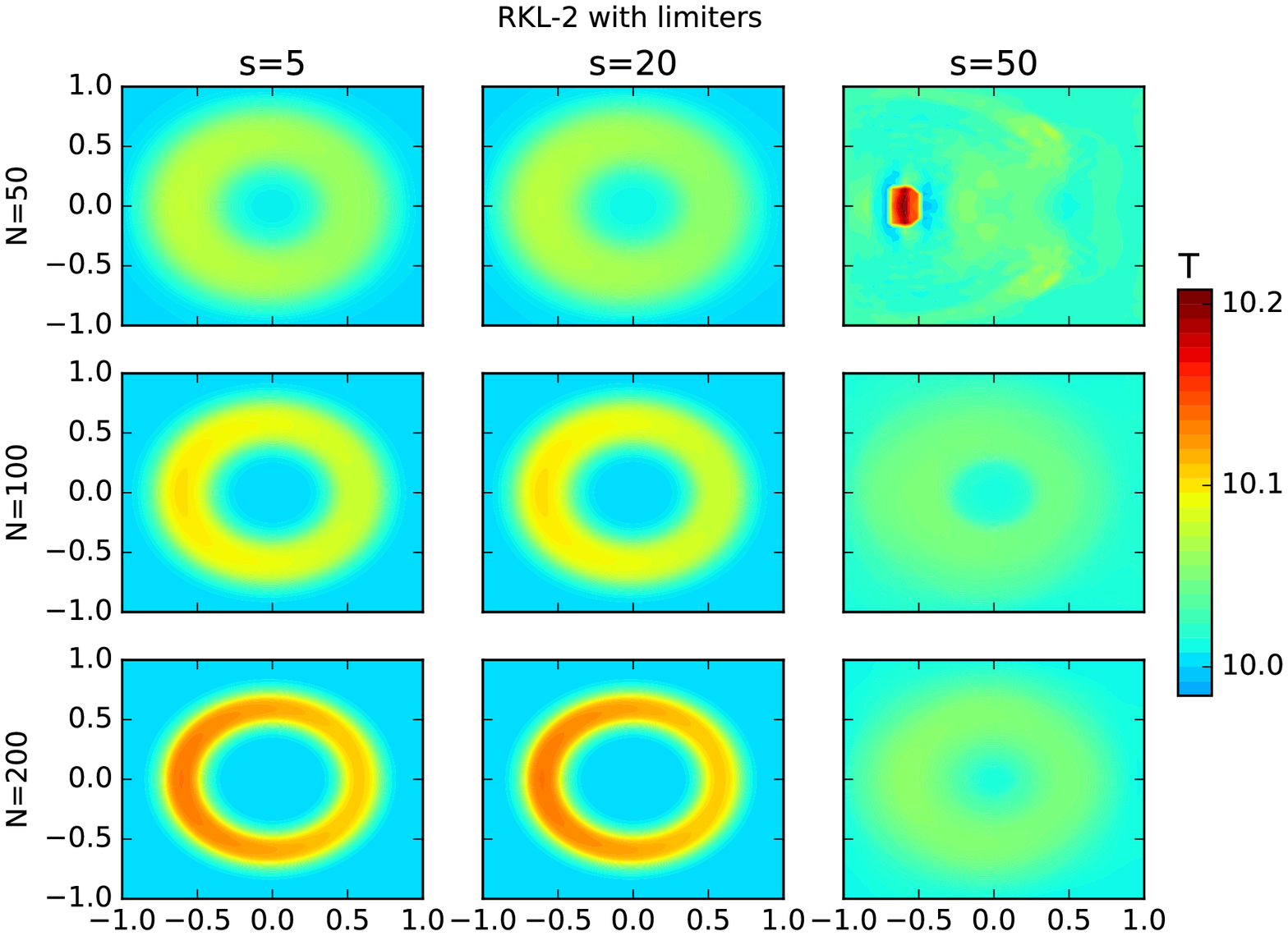}
  \caption{Temperature (in code units) contour plots for the ring diffusion test problem at $t=1$ evolved using RKL-2 STS without  (left panel) and with (right panel) limiters 
  for calculating the transverse temperature gradients in Eq. \ref{eq:qN_qT}. The grid resolution is $(50, 100, 200)$ and the number of RKL stages is $5, 20, 50$.
  Without limiters, the minimum temperature is always smaller than 10, the initial minimum. With limiters, the evolution is more diffusive and the temperature is below 
  10 only for $s=50$ and $N=50$.}
\label{fig:ring_diff}
\end{figure*}

 Figure \ref{fig:ring_diff} shows the temperature at
$t=1$ using RKL-2 scheme with different resolutions ($N=50,~100,~200$) and different number of STS stages ($s=20,~50,~100$); left (right) panel 
shows results without (with) limiters for interpolating transverse temperature gradients (see Appendix \ref{app:tr_lim}). For all cases without limiters the 
minimum temperature at $t=1$ is less than 10, the initial minimum temperature.\footnote{For the present test problem with a maximum and minimum temperature of 
12 and 10, the minimum temperature at later times is not negative but it will become negative if the initial minimum and maximum temperatures are say 0.1 and 10. 
A temperature ratio of 100 is commonplace in multiphase astrophysical flows, such as the interstellar medium. Once temperature becomes negative the MHD solver 
breaks down because of an imaginary sound speed.} 
With limiters, only $s=50$ and $N=50$ run shows a temperature
below 10. Therefore, monotonicity is better maintained with the use of limiters.  A comparison of left and right panels shows that the use of limiters introduces
large numerical diffusion perpendicular to field lines, especially for smaller resolution and larger number of stages (i.e., for smaller $N$ and larger $s$). A similar
behaviour is observed for other schemes such as AAG-STS, RKC-STS and RKL-1. For very large $\Delta t_{\rm h}/\Delta t_{\rm p}$, corresponding
to $s \sim \sqrt{\Delta t_{\rm h}/\Delta t_{\rm p}} \gtrsim 20$, it is better
not to use the limiters because of excessive transverse diffusion. (A very large $s$, although numerically stable, also leads to a loss of accuracy.)
However, limiters are necessary for preventing negative temperatures in presence 
of large gradients. For practical implementation we recommend the use of limiters only when temperature becomes smaller than a reasonable floor value in 
the anisotropic conduction step.

\subsection{MHD Tests}
\label{sec:supernova}
%
%
Unlike section \ref{sec:scalar_tests}, the test problems described in this section
solve the full set of MHD equations in presence of thermal conduction (Eqs. \ref{eq:mass}-\ref{eq:B}).
 These tests demonstrate the
coupling of the MHD hyperbolic conservation laws with
the parabolic update of  thermal conduction.  

\subsubsection{Supernova blast-wave}
\label{sec:mhdblast}
%

We consider an MHD blast wave in cylindrical co-ordinates with initial parameters similar to the $L1$ model of \citet{meyer2012second} but without radiative cooling. For this 2D axisymmetric test, we solve the standard set of ideal MHD equations, taking into account anisotropy in thermal conduction flux along with saturated conduction with $\phi=0.3$ (see Eqs. \ref{eq:Fc}, \ref{eq:Fclass} and \ref{eq:Fsat}).

The initial magnetic field of $0.3~\mu$G is oriented along the $z-$ axis.Energy of $10^{51} \rm erg$, equivalent to a supernova explosion, is injected in a region of spherical radius ($[r^2 + z^2]^{1/2}$) $7.8$\,pc around the origin of a $300$\,pc 
domain in  cylindrical geometry. A high pressure is set inside this spherical region (resolved with 10 zones) such that one-third of the supernova energy goes as thermal energy. The radial velocity of this hot material is set by equating the kinetic energy to the remaining two-third of energy.  
The temperature in the ambient medium is set to be $8000$ \,K and the initial density is set to be $0.7 m_u$ cm$^{-3}$ ($m_u$ is atomic mass unit) throughout the domain.  
Axisymmetric boundary conditions are imposed at $r = 0.0$ and initial conditions are imposed at the vertical boundary $z = 0.0$. The conditions at the outer boundaries in both $r-$ and $z-$ directions are imposed to be outflow.  

\begin{figure*}
\begin{center}
\includegraphics[width=2\columnwidth]{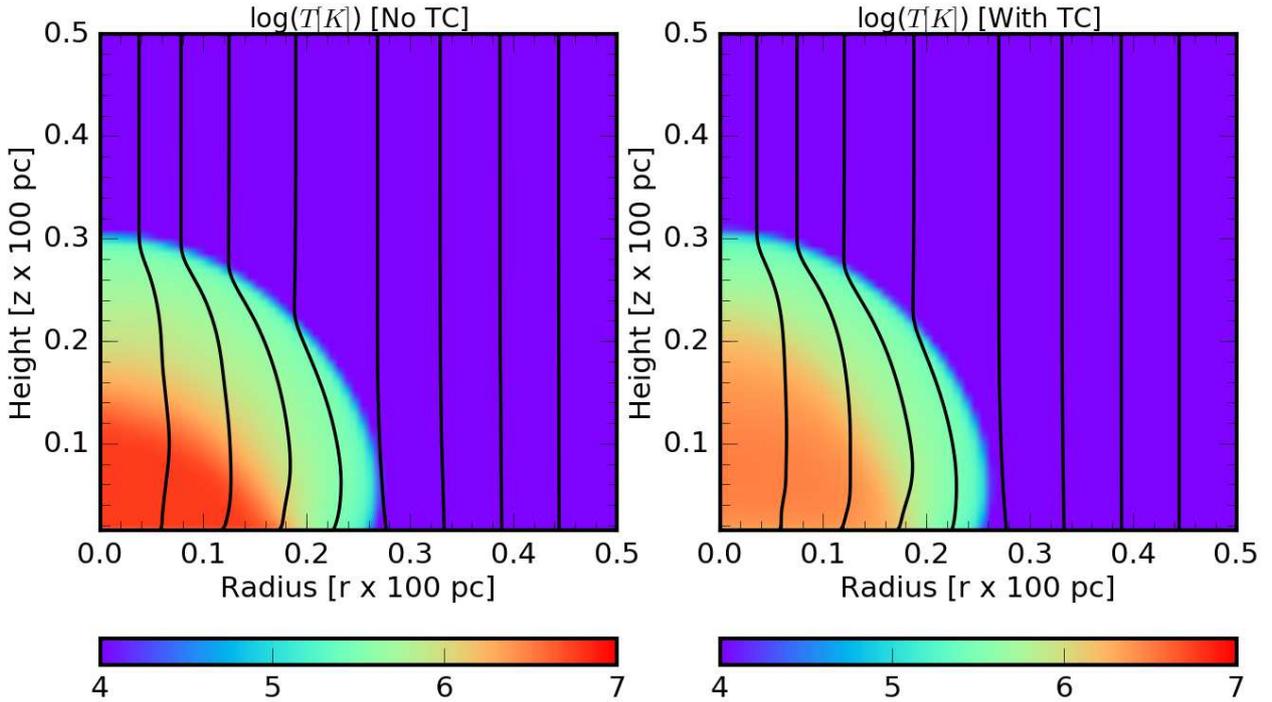}
\caption{Comparison of logarithmic value temperature (in K) for the \RefC{2D} magnetised blast wave test problem without (left panel) and with (right panel) anisotropic thermal conduction. The RKL-2 method is used for the run with conduction. The snapshot shown is at time $t = 0.1$ Myr. The black lines indicate the magnetic field lines.}
\label{fig:blast_anim}
\end{center}
\end{figure*}

Figure \ref{fig:blast_anim} shows the comparison of logarithmic value of temperature in Kelvin 
for runs with and without Thermal Conduction at time $t = 0.1$ Myr. 
The RKL runs with thermal conduction use Spitzer conduction along field lines and no conduction across the field lines.
The black lines shown in each panel depict the magnetic field lines. 
Since thermal conduction flux $\bm{F_{\rm c}}$ is only along the field lines, this anisotropy leads to an asymmetric expansion of the
inner hot bubble along $z-$ axis. The evolution of temperature and magnetic field lines in Figure \ref{fig:blast_anim} evidently shows that 
the diffusion in the transverse direction is suppressed for the run with anisotropic thermal conduction. While in absence of thermal conduction the outer shock maintains its spherical structure.

Further, we compare the execution times for this test using the
standard explicit time-stepping and the  RKL-2 method. This comparison for varying grid resolutions is listed in table
\ref{tab:blast_times}.  The RKL-2 method is faster than the explicit method by one order of magnitude for our highest resolution run with a grid resolution of $1536^2$.
 \begin{table}
   \centering
   \caption{A comparison of the execution time for the blast wave problem}
   \label{tab:blast_times}
   \begin{tabular}{lrr} 
     \hline
     \multirow{2}{*}{$N$} & \multicolumn{2}{c}{Wall-clock time $[{\rm s}]$}\\ 
     \cline{2-3} & explicit & RKL-2\\
     \hline
     $192$  & $292$ & $112$\\
     $384$ & $4448$ & $1276$\\
     $768$ & $62700$ & $11828$\\
     $1536$ & $1.11 \times 10^6$ & $1.45 \times 10^5$\\
     \hline
   \end{tabular}
 \end{table}
This observation is consistent with the fact that STS techniques like RKL-2 method provide an \textit{asymptotic gain} over standard explicit time-stepping methods as demonstrated in the middle panel of Fig. \ref{fig:gauss_diff_err1} for the 1D Gaussian diffusion test (Sec. \ref{sec:gaussian1d}).

\subsubsection{Local thermal instability}
\label{sec:LTI}
%
%
In this section we describe the results from 2-D MHD simulations of local thermal instability (TI) with anisotropic conduction (\citealt{2010ApJ...720..652S}).  
Our computational domain is a 40 kpc $\times$ 40 kpc periodic Cartesian box with a mean initial electron density of $n_e=0.1$ cm$^{-3}$ (mean mass per 
particle/electron, $\mu/\mu_e=0.62/1.17 m_p$) and a uniform 
temperature of 0.7 keV (typical cool cluster core parameters). Only classical thermal conduction with the conductivity given by the Spitzer value (Eq.
11 in \citealt{2010ApJ...720..652S}) is included; i.e., $\bm{F}_{\rm c} = \bm{F}_{\rm class}$ instead of Eq. \ref{eq:Fc}.
We use a tabulated cooling function corresponding to the plasma of a third
solar metallicity (\citealt{1993ApJS...88..253S}); the cooling function is set to zero for $T<10^6$ K. 
Following \citet{2010ApJ...720..652S}, we include a spatially uniform heating rate density which is equal to the average cooling rate density over the 
computational domain. With this, the computational box is in {\it global} thermal balance, and this mimics the observed rough thermal 
balance inferred in cool core clusters. Moreover, this setup shows the exponential linear growth of the local thermal instability.
With these parameters, the initial cooling time (which is approximately equal to the 
growth timescale of local TI; e.g., see Eq. 19c in \citealt{2012MNRAS.419.3319M}) is 0.095 Gyr.
A magnetic field of 5 $\mu$G is initialised at 45 degrees to the Cartesian box. 
Homogeneous, isotropic, isobaric random density perturbations are initialised to seed the local TI. The density perturbations 
($\delta \equiv [\rho-\rho_0]/\rho_0$) are given by
\begin{equation}
\delta(x,y) = \sum_{|k|=2,|l|=2}^{10} a_{k,l} \cos \left ( \frac{2 \pi (k x + l y )}{L} + \phi_{k,l} \right),
\end{equation}
where $k, l$ are mode labels, $a_{k,l} = 1.5 \times 10^{-3} r (k^2+l^2)^{-1/2}$ and $\phi_{k,l,m}=2 \pi r$ ($r$ is a random number uniformly distributed between 
-0.5 to 0.5, which is different for the amplitude and phase and for different modes), and $L=40$ kpc is the box size. These choices give a maximum over-density
amplitude ${\rm max} (\delta) \approx 0.003$.
The setup is very similar to but not identical as \citet{2010ApJ...720..652S}.

We run the local TI test problem using different methods for anisotropic thermal conduction: (a) {\it fully explicit}
evolution in which both the MHD and conduction modules are evolved using a time step $\Delta t_{\rm p}$ (Eq. \ref{eq:dt_exp};  this is typically shorter than 
the MHD CFL step $\Delta t_{\rm h}$; Eq. \ref{eq:dt_h}); (b) {\it sub-cycling} of conduction module in which MHD module uses $\Delta t_{\rm h}$ but the conduction
module is sub-cycled and applied $\Delta t_{\rm h}/\Delta t_{\rm p}$ times using a time step of $\Delta t_{\rm p}$; (c) MHD module is evolved 
using $\Delta t_{\rm h}$ and conduction module is evolved using  \textit{AAG-STS} with $\nu=0.01$ and $s \sim (\Delta t_{\rm h}/\Delta t_{\rm p})^{1/2}$ 
stages (see Eq. \ref{eq:AAG-STS}); 
and (d) MHD module is evolved using $\Delta t_{\rm h}$ and conduction module is evolved using RKL-2 with 
$s \sim  (\Delta t_{\rm h}/\Delta t_{\rm p})^{1/2}$ stages (see Eq. \ref{eq:RKL-s}). 
The wall-time taken for different methods to run the $512\times 512$ TI test problem \RefC{until} 0.87 
Gyr (9.16 cooling times) is listed in Table \ref{tab:TI}. All the methods use the monotonized-centered (MC) limiter to calculate the transverse terms in the 
anisotropic heat flux ($F_x^T$ in Eq. \ref{eq:qT} and analogous expression for $F_y^T$). 
If we do not use limiters for interpolating transverse temperature 
gradients (and instead use simple averaging), all the different runs (a-d) blow up at some point in nonlinear evolution due to negative temperature somewhere
in the computational domain. This test highlights the importance of using limiters for robustness in presence of large temperature gradients.

\begin{table}
\caption{Wall-time for the local TI test run till 0.87 Gyr }
\resizebox{0.48 \textwidth}{!}{%
\begin{tabular}{c c c c }
\hline
explicit & sub-cycling & AAG-STS$^\dag$ & RKL-2 \\
\hline
6 h 1 m 29 s & 43 m 45 s & 9 m & 16 m 6 s  \\
\hline
\end{tabular}}
\label{tab:TI}
The grid resolution is $512\times 512$.
$^\dag$ AAG-STS run crashes at 0.87 Gyr, but others do not.
\end{table}

We note that the AAG-STS run crashes at 0.87 Gyr even with limiters because of a large $C_p = \Delta t_{\rm h}/2\Delta t_{\rm p}$ (see Fig. \ref{fig:gauss_diff_err2}). 
Other runs could go for much longer without numerical problems. 
Both RKL-2 and AAG-STS methods clearly show a significant speed-up relative to sub-cycling and explicit methods.
\RefC{On comparing the STS methods, we see that AAG-STS is faster than RKL-2 due to a smaller number of computations per cycle.
However, the first-order accurate AAG-STS scheme exhibits unstable behavior while RKL-2 maintains stability during the entire integration as described below.}

\begin{figure*}
\centering{
\includegraphics[scale=0.75]{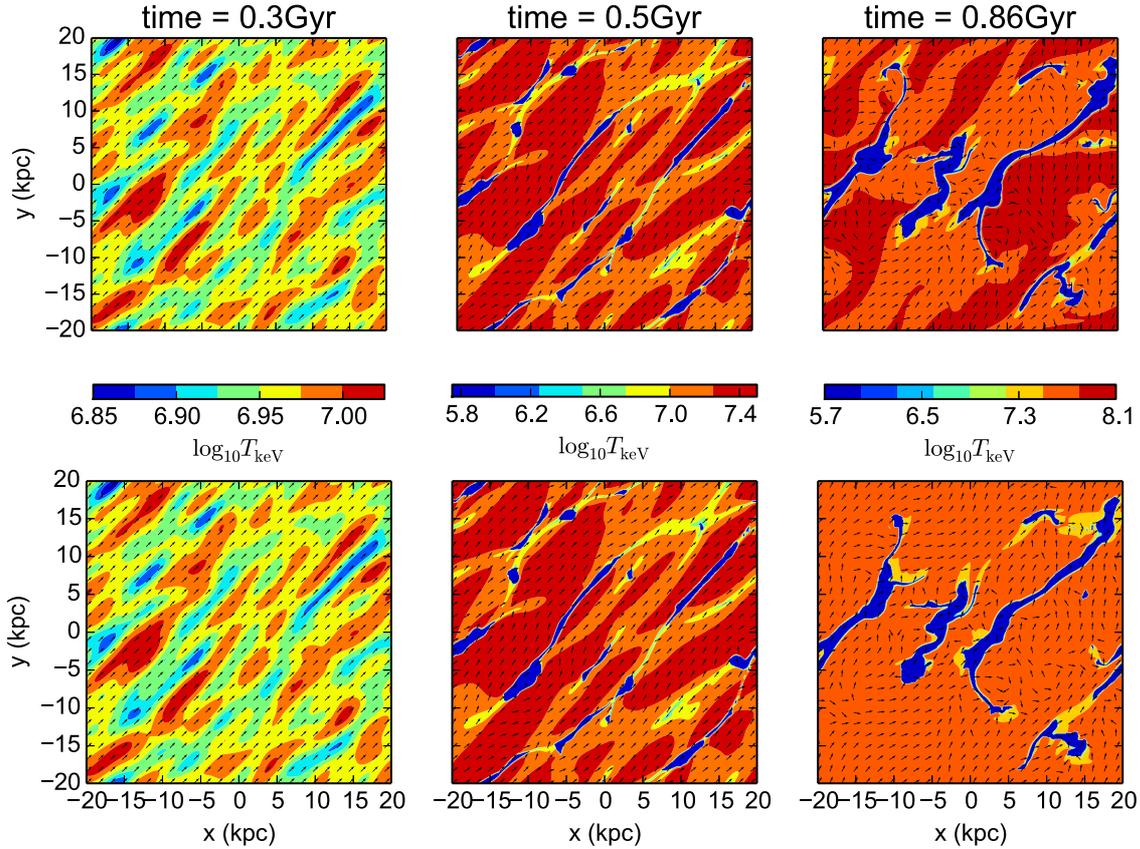}
}
\caption{Contour plots of log$_{10}$ temperature (keV) for RKL-2 (top panels) and AAG-STS (bottom panels) runs at linear (0.30 Gyr) and nonlinear 
(0.45 Gyr and 0.86 Gyr) stages of the instability. The arrows shows the local magnetic field unit vector. The non-linear stage starts at around 
0.45 Gyr and the magnetic field lines no longer remain aligned at 45$^0$ to the x-axis. The evolution of TI differs substantially with the two 
methods at late times in the non-linear phase. The AAG-STS run crashes at 0.87 Gyr.
}
\label{fig:COMPARE}
\end{figure*} 

Figure \ref{fig:COMPARE} shows temperature snapshots at different stages of TI evolution using RKL-2 (top panels) and AAG-STS (bottom panel) schemes. 
In the linear stage all schemes give a similar evolution. 
In the saturated nonlinear  state AAG-STS and RKL-1 start to deviate quantitatively from each other. Unlike AAG-STS, the RKL-2 temperature 
snapshots at 0.86 Gyr are very similar to the ones obtained using explicit and sub-cycling methods (not shown in Fig. \ref{fig:COMPARE}); this time is very 
close to the time when the AAG-STS run blows up (at 0.87 Gyr) due to numerical instability. 

\begin{figure*}
\centering{
\includegraphics[scale=0.56]{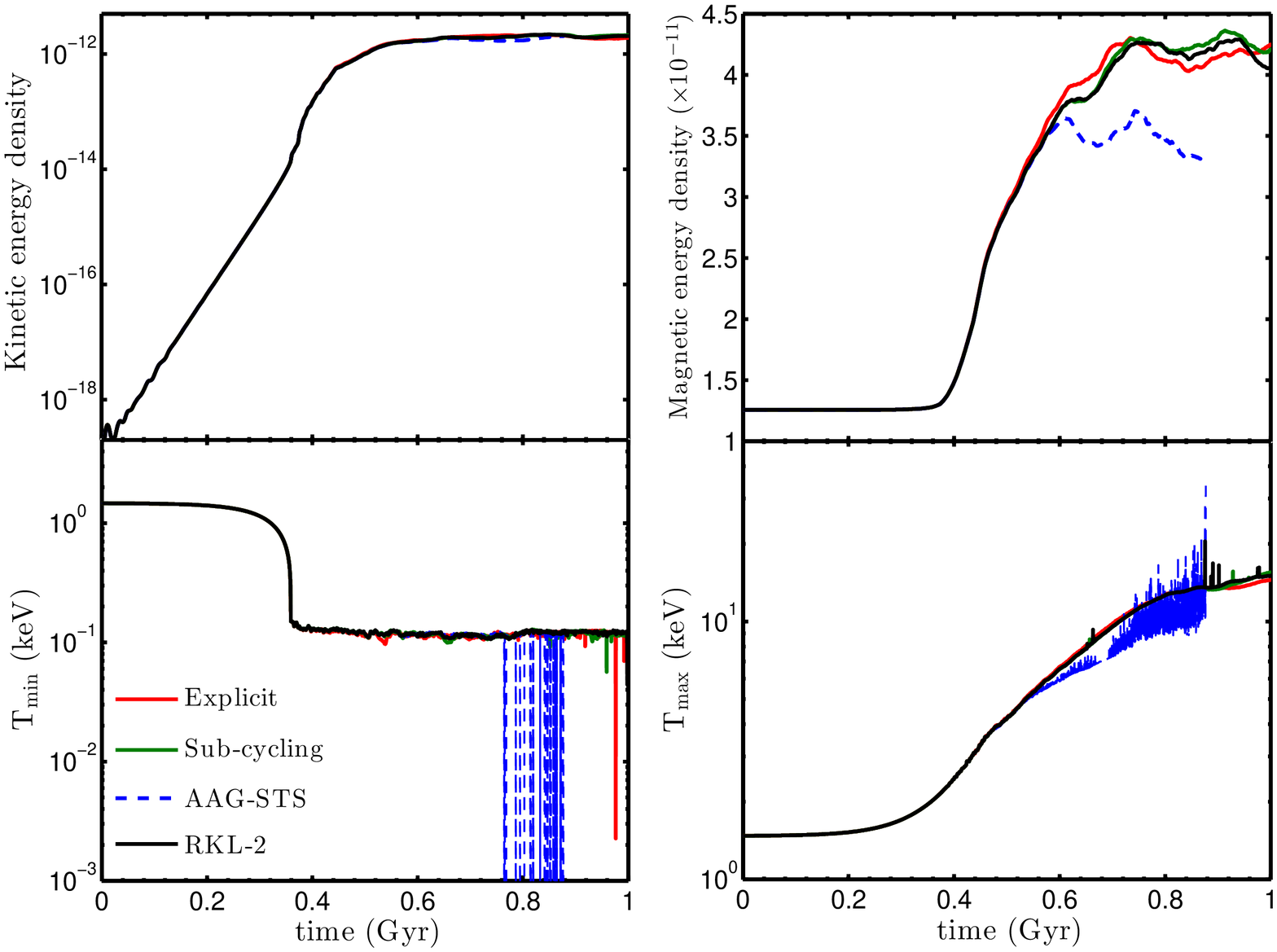}
}
\caption{The evolution of various quantities like kinetic energy density [ergs cm$^{-3}$], magnetic energy density [ergs cm$^{-3}$], the minimum and maximum temperature in the computational domain ($T_{\rm min}$ and $T_{\rm max}$) with time using different methods for anisotropic conduction \RefC{to study the local TI} . Explicit (red solid line), sub-cycling (green solid line), and RKL-2 (black solid line)  methods show similar time evolution but AAG-STS (blue dashed line) starts to deviate in the non-linear stage. Notice the numerous spikes in $T_{\rm max}$ and $T_{\rm min}$ for AAG-STS, which are symptoms of numerical instability which blows up the code at 0.87 Gyr. We impose a numerical temperature floor when the temperature becomes negative.
}
\label{fig:time_evol}
\end{figure*}

Figure \ref{fig:time_evol} shows the time evolution of various quantities as a function of time using different methods for anisotropic thermal conduction. 
Top left panel of Figure \ref{fig:time_evol} shows that the evolution of kinetic energy in the box is similar for all the runs. The top right panel 
shows  the average magnetic energy evolution. Here, AAG-STS deviates from the other runs in the non-linear stage. Similar deviations are seen in AAG-STS 
for $T_{\rm min}$ and $T_{\rm max}$ (minimum and maximum temperature in the computational domain), but not for RKL-2. The numerical fragility of AAG-STS due to an imperfect stability parameter $\nu$ is evident 
from spikes in  $T_{\rm min}$ and $T_{\rm max}$. The AAG-STS run blows up at 0.87 Gyr due to the numerical instability of AAG-STS for a large number of 
sub-stages.

\section{Parallel scaling of RKL conduction}
\label{sec:par_scaling}
%
%

In this section, we demonstrate the parallel scaling of isotropic and anisotropic thermal conduction in PLUTO code.
For this purpose, we use the blast wave test problem in three dimensions on a Cartesian grid with size $L = 30$ pc.  
To set up a blast wave we initialise an ambient static medium with density $\rho_0$ and pressure $p_0$. 
A blast region with high density and pressure is set within a spherical radius $r_0$ around the origin. 
The density and pressure in this region is a factor 10 and 1000 times larger than the ambient medium respectively. 
We smooth the transition between the blast region and the ambient medium using a smoothing function as follows
\begin{eqnarray}
\rho &=& \rho_0 (1 + 9 \Lambda[r]) \nonumber \\
p &=& p_0 (1 + 999 \Lambda[r])/\gamma
\end{eqnarray}
where, $\Lambda[r]=(1 + \exp( [r- r_0]/[0.1~{\rm pc}]))^{-1}$ is a smoothing function, 
$r=(x^2+y^2+z^2)^{1/2}$ is the spherical radius, $p_0=2.295\times10^{-9}$ dyn cm$^{-2}$,  
$\rho_0=2.1\times10^{-23}$ g cm$^{-3}$, $r_0=1.0$ pc, and $\gamma=5/3$. 
Additionally, for MHD runs the initial magnetic field is along $y$ axis, $\vec{B} = B_0 \hat{\vec{y}}$ where $B_0  = \sqrt{2 p/\beta}$ with $\beta = 1$.
Thus, the field strength inside the high pressure blast region is about 30 times larger than the ambient medium.
We use Spitzer conduction along field lines and no conduction across the field lines for the MHD run. The HD run uses isotropic
Spitzer conduction. For comparison, we also include a HD run without conduction.

We carried out strong scaling studies using a $512^3$ Cartesian grid
on CPU-only nodes of IISc Cray XC40 cluster SahasraT\footnote{\textit{http://www.serc.iisc.in/facilities/cray-xc40-named-as-sahasrat/}}.
\RefCC{Unlike the 2-D blast wave test problem in section \ref{sec:mhdblast}, we choose a 3-D blast wave problem for testing parallel strong scaling on more than ten thousand processors. In strong scaling, the problem size remains fixed but the number of processors is increased progressively. A large enough problem size is needed (that is why 3-D instead of 2-D test problem is chosen) for communication not to dominate over computation even on largest number of processors.
}
We run the blast wave test problem for $t = 0.5$ kyr on processors ranging from 256 to 16384. 
On average, the number of sub-steps $s$ ranges between 5-12 for the HD run and between 3-8 for MHD. 

\begin{figure*}
\includegraphics[width=2\columnwidth]{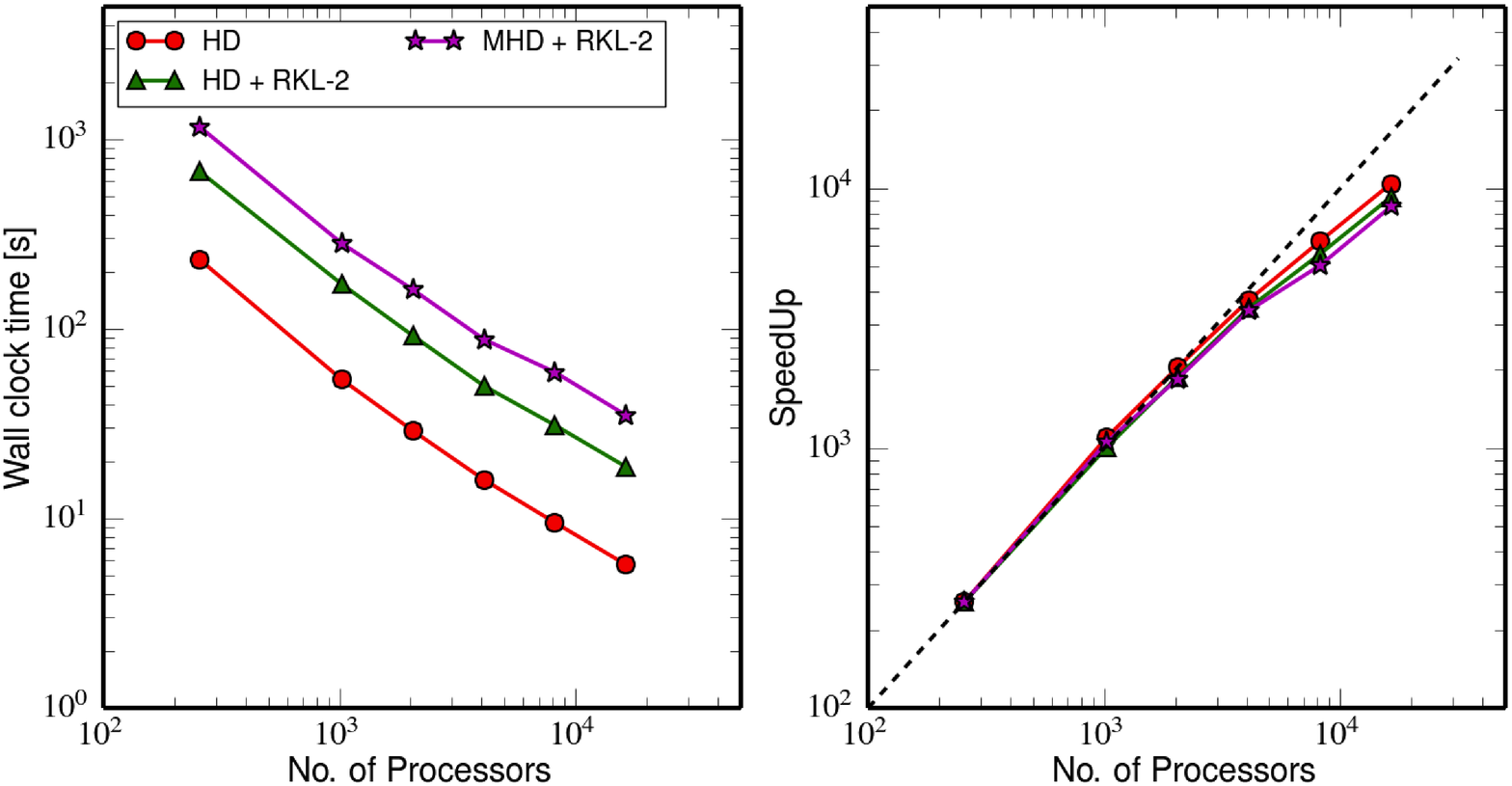}
\caption{Strong scaling test for the 3D blast wave test with 512$^{3}$ cells. The \textit{left panel}
  shows the wall clock time for the simulations run till $t = 0.5$ kyr using different number of processors. 
  The \textit{right panel} shows the comparison of numerical speed-up with the ideal scaling (black dashed line). 
  Scaling study using pure HD (\textit{red circles}), HD+RKL (\textit{green triangles}) and MHD+RKL (\textit{magneta stars})
  shows a close to perfect scaling up to 10$^4$ processors.    
}
\label{fig:rkl2_parascale}
\end{figure*}

The left panel of Figure \ref{fig:rkl2_parascale} shows the wall clock time as a function of processor count for our strong scaling studies. 
The wall clock time behaves as expected, with the MHD-RKL run taking longer than the HD-RKL run. Both runs with conduction show 
close to the inverse linear scaling with the processor count. The right panel of Figure \ref{fig:rkl2_parascale} shows that the speed-up
improves with increasing core count, but drops to 80\% for more than $10^4$ processors.
The same trend is also seen for pure HD run, indicating that the addition of thermal conduction has not resulted in any performance degradation.
Our strong scaling tests show that RKL methods yield scalable algorithm on PetaFlop facilities.

\RefC{
This feature is usually not shared by implicit methods which, as already mentioned in Sec. \ref{sec:implmeth}, require inverting large sparse matrices, an operation difficult to achieve efficiently on multi-core systems \citep{Botchev:2001}.
Nevertheless, the debate on which approach can be more efficient in massive parallel computations has yet to be settled.
In the work by \cite{Pakmor:2016}, for instance, a semi-implicit scheme is employed to solve the cosmic-ray transport equation (coupled to the MHD equations) on unstructured moving mesh in the context of galaxy dynamics.
Their semi-implicit solver requires solving a single linear system of equations per timestep and weak scaling tests claim good parallel efficiency, up to 480 cores.
However, more complex scenarios including nonlinear systems of equations (e.g Ohmic diffusion, ambipolar diffusion, etc.) may introduce additional complexities making the applicability of implicit scheme less efficient.
Although a comparison with implicit methods is outside the scope of this paper, we point out to the recent work by \cite{Caplan:2017}, where explicit STS schemes are compared with implicit Krylov solvers in the context of magnetised solar corona.
The left panels of Fig. 11 in their paper shows the superiority of explicit schemes over implicit ones for their choice of problem and model.   
The comparison also shows that the strong parallel scaling of implicit methods saturates and deviates sharply from ideal behavior at large number of cores ($>1000 $). 
}

\RefC{
The explicit RKL schemes presented in this work confirm this prediction (the scaling performance with explicit RKL method shown in Fig. \ref{fig:rkl2_parascale} is able to achieve a high efficiency of 80\% even up to 1000 cores) and can be naturally extended  to more complex systems of equations with with minimal modifications.
}

\section{Conclusions}
\label{sec:conclusions}
%
%
%
%

In this paper we have discussed various numerical methods implementing anisotropic thermal conduction coupled with the standard set of 
MHD equations. In particular, we have described 
the second order (in time) accurate Runge-Kutta Legendre (RKL-2)  super-time stepping (STS) method implemented in PLUTO code. 
We have then compared these numerical methods on simple test problems, and also on astrophysical test problems like blast wave and  thermal instability in which 
MHD evolution is coupled with conduction.
The major conclusions of our paper are:
\begin{enumerate}
\item Using the 1-D Gaussian diffusion test, we show that the RKL-2 scheme is second order accurate in time, in comparison to the standard AAG-STS (\citealt{alexiades1996super}) scheme which is first order.
\item Super-time stepping schemes based on Chebyshev polynomials such as  AAG-STS require an ad-hoc damping parameter ($\nu$), which has to be big enough 
for numerical stability. Figure \ref{fig:gauss_diff_err2} shows that AAG-STS becomes unstable for number of substages $s \gtrsim$ few 10s. Moreover, for 
$2 s \sqrt{\nu} \gtrsim 1$ there is no real speed-up using AAG-STS (see Eq. \ref{eq:STS-limit}). In absence of such a parameter and with sufficient inherent damping, 
RKL-STS schemes are more robust and can better exploit the super-time stepping strategy (see the right panel of Fig. \ref{fig:gauss_diff_err1}). The robustness of RKL-STS
is also useful for implementing saturated conduction (e.g., see Fig. \ref{fig:sawtooth_comp}).
\item The Cartesian ring diffusion test problem (Fig. \ref{fig:ring_diff}) shows that all STS schemes break down if we use a large number of stages with low resolution. Moreover,
the use of limiters to interpolate the transverse temperature gradient leads to somewhat larger perpendicular diffusion. Limiters do help prevent non-monotonicity of temperature
in presence of temperature discontinuities. For practical implementation of anisotropic diffusion, limiters need not be used as a default option, but may be recommended at locations and times at which the temperature falls below a floor value. The local thermal instability test problem in section \ref{sec:LTI} demonstrates the utility of limiters in presence of temperature discontinuities.
\item As demonstrated in Tables \ref{tab:scaling}, \ref{tab:blast_times} \& \ref{tab:TI}, and also indicated by the middle panel of Figure \ref{fig:gauss_diff_err1}, 
the run-time with super-time stepping is substantially shorter than with explicit update or even sub-cycling.
\item Last, but perhaps most importantly as shown in Figure \ref{fig:rkl2_parascale}, the explicit schemes such as RKL-STS shows 
an \RefC{ excellent scaling (efficiency of $\sim 80\%$ up to 10$^4$ processors)} on modern distributed PetaScale supercomputers. 
\end{enumerate} 
  
\section*{Acknowledgments}
The authors would sincerely like to thank the referee for the valuable comments 
which have played a significant role in improving the paper. 
BV would like to thank the support provided by University of Torino and also would like to acknowledge
the hospitality of IISc during the research visit in 2016. 
This work is partly supported by the DST-India grant no. Sr/S2/HEP-048/2012 and an India-Israel
joint research grant (6-10/2014[IC]). We thank the SERC-IISc staff for facilitating our use of SahasraT 
cluster for the parallel scaling runs; only with their intervention we could finish our runs in a reasonable time.
PS acknowledges the hospitality of KITP where this paper was completed. 
This research was supported in part by the National Science Foundation under Grant No. NSF PHY-1125915.

\appendix
\section{\RefC{Illustration of RKL-2 scheme}}
\label{sec:exprkl2}
This Appendix is closely based on \citet{meyer2012second,meyer2014stabilized} and is included here 
for completeness.
Second order accuracy for the RKL-2 scheme can be achieved by matching the first three terms in Eqs. (\ref{eq:tayexp}) \& (\ref{eq:stabpara}); 
i.e., by imposing $R_{s}(0)$ = 1, $R_{s}^{'}(0)$ = 1 and $R_{s}^{''}(0)$ = 1. 
These three requirements can be used to estimate the values of three coefficients in Eq.(\ref{eq:stabpara}) as,
\begin{eqnarray}
  a_{s}   &=&  1 - b_{s} \\
  b_{s}   &=&  \frac{\legPpp{s}{1}}{(\legPp{s}{1})^2} = \frac{s^2 + s - 2}{2s(s+1)}  \,\,\,\,\,\,\,s \ge 2 \\
  w_1     &=&  \frac{\legPp{s}{1}}{\legPpp{s}{1}} = 
  \frac{4}{s^2 + s - 2} \,\,\,\,\,\,\,s \ge 2 
\end{eqnarray}
For $s < 2$, we can choose $b_0 = b_1 = b_2 = 1/3$ \citep{meyer2014stabilized}. 
Here we have used various properties of the Legendre polynomials  and their derivatives, i.e.,
\begin{eqnarray}
 \legP{s}{1} &=& 1  \\
 \legPp{s}{1} &=& \frac{s(s+1)}{2} \\
 \legPpp{s}{1} &=& \left(\frac{s^2 + s - 2}{4}\right) \legPp{s}{1}
\end{eqnarray}
Like Eq. (\ref{eq:stabpara}) for the stability polynomial at the end of $s$ stages, the stability polynomial till $j$ stages is chosen to be
$a_j + b_j \legP{j}{1+w_1 (\tau \lambda)}$.

The RKL-2 scheme with $s$ stages can be expressed as,
\begin{eqnarray} 
\label{eq:rkl2scheme}
  Y_0 &=& u(t_0) \nonumber\\
  Y_1 &=& Y_0 + \tilde{\mu}_{1} \tau \matm Y_0 \nonumber \\
  Y_j &=& \mu_{j} Y_{j-1} + \nu_{j} Y_{j-2} + (1 - \mu_{j} - \nu_{j}) Y_0 \nonumber\\
  && + \tilde{\mu}_{j} \tau \matm Y_{j-1} + \tilde{\gamma}_{j} \tau \matm Y_0  \hspace{1.6cm} 2 \le j \le s \nonumber\\
   u(t_0 + \tau) &=& Y_s
\end{eqnarray}
where the coefficients,  $\mu_{j}$, $\tilde{\mu}_{1}$, $\tilde{\mu}_{j}$, $\nu_j$ and $\tilde{\gamma}_{j}$ can be obtained using the recursion relation for the Legendre polynomials (see Eq.(\ref{eq:recLegP})) and rearrangement of terms. 
The expressions for these coefficients are given by,
\begin{eqnarray}
 \mu_j &=& \frac{2j - 1}{j} \frac{b_j}{b_{j-1}} \\
 \tilde{\mu}_{1} &=& b_1 w_1\\
 \tilde{\mu}_{j} &=& \mu_j w_1 \\
  \nu_{j}   &=& -\frac{j-1}{j} \frac{b_j}{b_{j-2}}\\
 \tilde{\gamma}_j &=& -a_{j-1} \tilde{\mu}_j
\end{eqnarray}

To get a better sense of the scheme, we explicitly list the value of the above coefficients for a representative small value of $s=3$. For this three stage scheme we have,

\begin{eqnarray}
 \mu_{2} &=& \frac{3}{2};\,\,\, \mu_{3} = \frac{25}{12} \nonumber\\
 \tilde{\mu}_1 &=& \frac{2}{15}; \,\,\, \tilde{\mu}_{2} = \frac{3}{2}w_1 = \frac{3}{5}; \,\,\, \tilde{\mu}_{3} = \frac{25}{12}w_1 = \frac{5}{6} \nonumber \\
 \nu_{2} &=& -\frac{1}{2} ; \,\,\, \nu_{3} = -\frac{2}{3}\frac{b_3}{b_1} = -\frac{5}{6} \nonumber\\
 \tilde{\gamma}_{2} &=& (b_1 - 1)\tilde{\mu}_{2} = -\frac{2}{5}; \,\,\, \tilde{\gamma}_{3} = (b_2 -1)\tilde{\mu}_{3} = -\frac{5}{9} \nonumber
\end{eqnarray}

On substituting the above values of various coefficients for $s=3$ in Eq.(\ref{eq:rkl2scheme}) we get;
\begin{eqnarray}
 Y_0 &=& u(t_0) \\
  Y_1 &=& Y_0 + \frac{2}{15} \tau \matm Y_0\\
      &=& Y_0 + \frac{2}{15} \tau \lambda Y_0 \nonumber\\
      &=& \left(1 + \frac{2}{15} (\tau \lambda)\right) Y_0 \nonumber\\
      &=& \left(\frac{2}{3} + \frac{1}{3} \legP{1}{1 + \frac{2}{5}(\tau\lambda)}\right) Y_0 \nonumber \\
      &=& \left(a_1 + b_1 \legP{1}{1 + w_1 (\tau\lambda)}\right) Y_0 = R_1(\tau \lambda) Y_0 \nonumber \\
 Y_2  &=& \frac{3}{2} Y_1 - \frac{1}{2} Y_0 + \frac{3}{5} \tau \matm Y_1 -\frac{2}{5} \tau \matm Y_0 \\
  &=& \left(\frac{3}{2} + \frac{3}{5} (\tau \lambda)\right)Y_1 - \left(\frac{1}{2} + \frac{2}{5} (\tau \lambda)\right) Y_0   \nonumber  \\
  &=& \left(1 + \frac{2}{5}(\tau \lambda) + \frac{2}{25}(\tau\lambda)^2\right)Y_0 \nonumber \\
  &=& \left(\frac{2}{3} + \frac{1}{3}\legP{2}{1 + \frac{2}{5}(\tau\lambda)}\right) Y_0 = R_2(\tau\lambda) Y_0 \nonumber \\
 Y_3  &=& \frac{25}{12} Y_{2} -\frac{5}{6} Y_1 - \frac{1}{4} Y_0 + \frac{5}{6} \tau \matm Y_2 -\frac{5}{9} \tau \matm Y_0 \nonumber\\
 &=&  \left(1 + (\tau \lambda) + \frac{1}{2} (\tau \lambda)^2 + \frac{1}{15} (\tau\lambda)^3 \right)Y_0 \label{eq:comtay} \\
 &=& \left(\frac{7}{12} + \frac{5}{12}\legP{3}{1 + \frac{2}{5}(\tau\lambda)}\right)Y_0 = R_3(\tau \lambda) Y_0\nonumber \\
 u(t_0 + \tau) &=& Y_3
\end{eqnarray}
where $\lambda$ is the eigenvalue of parabolic operator matrix $\matm$. Note that the first three terms of final solution of $Y_3$ (Eq. \ref{eq:comtay}) match the Taylor expansion of the exponential function (Eq.(\ref{eq:tayexp}). Thus this scheme is second order-accurate with a leading order in error given by the fourth term, i.e., $(\tau \lambda)^3/15$. 

Finally to estimate the value of the super-step $\tau$, we should have $|R_3(\tau\lambda)| \le 1$. The stability condition will be satisfied only if the
argument of LP is $\geq -1$ (note that $\lambda$ is non-positive),  i.e., $-5 \le \tau \lambda \le 0$. While the similar stability condition for one step Euler-scheme requires $-2 \le \tau \lambda \le 0$. This clearly shows the $s=3$ stage RKL-2 method allows to choose a large time-step as compared to the Euler scheme. Higher values of sub-steps $s$, will result in a wider range of stable time steps giving an obvious advantage over standard explicit schemes.
 
\section{Limiting transverse temperature gradients for monotonicity}
\label{app:tr_lim}

Simple finite differencing of Eq. \ref{eq:scalar_diffusion} accentuates temperature extrema in presence of large temperature gradients expected
naturally in astrophysical plasmas
(e.g., \citealt{sharma2007preserving}). The heat flux can be decomposed into normal and transverse components (Eqs. \ref{eq:qN}, \ref{eq:qT}). 
Expressing the heat flux \sout{,} $\bm{F}_{\rm c}$ with components normal and transverse to the field lines 
we obtain the following explicit formulation of Eq. \ref{eq:scalar_diffusion} in 2D Cartesian geometry (generalisation to 3-D and non-Cartesian coordinates is straightforward),
\begin{equation}
\label{eq:qN_qT}
\frac{3}{2} n k_B \frac{\partial T}{\partial t} = - \frac{\partial }{\partial x} (F_{\rm c, x}^N + F_{\rm c, x}^T) - \frac{\partial }{\partial y} (F_{\rm c, y}^N + F_{\rm c, y}^T) .
\end{equation}

For simplicity we consider the classical limit (i.e., ignore saturated conduction). On comparing Eq. \ref{eq:qN_qT} with Eq. \ref{eq:Fclass}, we get
\begin{eqnarray}
\label{eq:qN}
F_{\rm c, x}^N &=& - \kappa_\parallel b_x^2 \frac{\partial T}{\partial x}, \\
\label{eq:qT}
F_{\rm c, x}^T &=& - \kappa_\parallel b_x b_y \frac{\partial T}{\partial y},
\end{eqnarray}
as the normal and transverse 
components of the heat flux (analogous expressions can be constructed for $F_{\rm c, y}^N$ and $F_{\rm c, y}^T$).

The normal component of the heat flux is naturally face-centred and always carries heat from higher to lower temperatures. However, the transverse 
temperature gradients are not naturally located at the faces, and need to be interpolated there. As the transverse flux can have any 
sign (Eq. \ref{eq:qT}), without special treatment, this component can lead to negative temperatures in regions with large temperature gradients (this also applies 
to saturated conduction in Eqs. \ref{eq:Fc} \& \ref{eq:Fsat}; therefore, the transverse temperature gradient required to evaluate ${\rm sgn}[\hvec{b} \cdot \nabla T$]
for saturated flux should also use limiters for robustness). 
\citet{sharma2007preserving} introduced limiters (similar to those used in the reconstruction step in finite volume methods; 
\citealt{leveque2002finite}) to interpolate the transverse temperature gradients at the cell faces and showed that the resulting explicit scheme 
preserves temperature extrema. 

\bibliographystyle{mnras}

\end{document}